\documentclass{jfm}

\usepackage{graphicx}
\usepackage{newtxtext}
\usepackage{newtxmath}
\usepackage{natbib}
\usepackage{hyperref}
\usepackage{enumitem}
\usepackage{comment}
\hypersetup{
    colorlinks = true,
    urlcolor   = blue,
    citecolor  = black,
}

\newcommand{\RomanNumeralCaps}[1]
\linenumbers
\usepackage{amsmath,amssymb}
\usepackage{amssymb}
\DeclareMathOperator{\sgn}{sgn}
\DeclareMathOperator{\Ma}{\mathcal{M}}
\DeclareMathOperator{\B}{\mathcal{B}}

\shorttitle{Surfactant amplifies yield-stress effects in capillary instability}
\shortauthor{J. D. Shemilt, A. Horsley, O. E. Jensen, A. B. Thompson and C. A. Whitfield}

\title{Surfactant amplifies yield-stress effects in the capillary instability of a film coating a tube}

\author{James D. Shemilt\aff{1}
  \corresp{\email{james.shemilt@manchester.ac.uk}},
  Alexander Horsley\aff{2},
  Oliver E. Jensen\aff{1},
  Alice B. Thompson\aff{1}
 \and Carl A. Whitfield\aff{1,2}}

\affiliation{\aff{1}Department of Mathematics, University of Manchester, Oxford Road, Manchester M13 9PL, UK
\aff{2}Division of Immunology, Immunity to Infection and Respiratory Medicine, University of Manchester, Oxford Road M13 9PL, UK}

\begin{document}
\maketitle

\begin{abstract}
To assess how the presence of surfactant in lung airways alters the flow of mucus that leads to plug formation and airway closure, we investigate the effect of insoluble surfactant on the instability of a viscoplastic liquid coating the interior of a cylindrical tube. Evolution equations for the layer thickness using thin-film and long-wave approximations are derived that incorporate yield-stress effects and capillary and Marangoni forces. Using numerical simulations and asymptotic analysis of the thin-film system, we quantify how the presence of surfactant slows growth of the Rayleigh-Plateau instability, increases the size of initial perturbation required to trigger instability and decreases the final peak height of the layer. When the surfactant strength is large, the thin-film dynamics coincide with the dynamics of a surfactant-free layer but with time slowed by a factor of four and the capillary Bingham number, a parameter proportional to the yield stress, exactly doubled. By solving the long-wave equations numerically, we quantify how increasing surfactant strength can increase the critical layer thickness for plug formation to occur and delay plugging. The previously established effect of the yield stress in suppressing plug formation [Shemilt et al., \textit{J. Fluid Mech.}, 2022, vol. 944, A22] is shown to be amplified by introducing surfactant. We discuss the implications of these results for understanding the impact of surfactant deficiency and increased mucus yield stress in obstructive lung diseases. 
\end{abstract}



\section{\label{sec:intro}Introduction}

Pulmonary surfactant plays a crucial role in the healthy function of the lungs. By lowering surface tension at the interface between air and the liquid that lines lung airways, surfactant helps to prevent unwanted collapse of small airways and alveoli during breathing \citep{milad_2021_revisiting}. Surfactant deficiency is likely to contribute to the increased prevalence of airway obstructions in diseases such as asthma \citep{hohlfield_2001_asthma} and cystic fibrosis \citep{tiddens_cystic_2010}. Mucus, the main component of the airway surface liquid, is a complex fluid exhibiting properties such as viscoelasticity, shear-thinning and a yield stress \citep{hill_2022_mucus}. In various obstructive diseases, and particularly in cystic fibrosis, the mucus typically has altered rheology, including a significantly increased yield stress compared to mucus in healthy lungs \citep{patarin_rheological_2020}. This provides motivation for this study into the effect of insoluble surfactant on the instability of a viscoplastic liquid coating the interior of a cylindrical tube, which is a simple model for the flow of mucus that can lead to airway closure. Additional applications can be found in related interfacial flows of viscoplastic fluids from engineering and industry, where surfactants may also be present \citep{mitsoulis_2007_review,glasser_2019_tuning,ahmadikhasmi_2020_impact}.

The instability of a liquid film coating the interior of a cylindrical tube has been widely studied in the case that the liquid is Newtonian. When the volume of fluid in the layer is small, annular liquid collars will form on the tube wall, and when the volume is large enough, liquid plugs form in the tube \citep{everett_model_1972}. \citet{hammond_nonlinear_1983} derived an evolution equation for the motion of a thin layer, and presented numerical and late-time asymptotic solutions showing annular collars forming and fluid slowly draining out of thin regions between them. This thin-film theory was then extended to describe the motion of thick films by \citet{gauglitz_extended_1988}, who deduced numerically that an approximate minimum thickness of $12\%$ of the tube radius is required for plug formation to occur. 

\citet{otis_effect_1990} were the first to extend the theory to include the effect of insoluble surfactant at the air-liquid interface. They found that growth of the instability and plug formation is delayed by surfactant. Moreover, if the surfactant is strong then Marangoni forces effectively immobilise the interface, increasing the time scale for the evolution of the layer by a factor of four compared to when there is no surfactant. This factor of four decrease in the growth rate had been previously identified by \cite{Carroll_1974_effect} in the related instability of a thin liquid layer coating the exterior of a cylindrical filament. Results from the thick-film model of \citet{ogrosky_linear_2021} suggest that, whilst this factor is very close to four for layers with thicknesses close to the critical value for plug formation, it may be increased for thicker layers. \citet{halpern_surfactant_1993} investigated the effect of surfactant on the evolution of a layer coating the interior of an elastic tube, and also found slowing of the dynamics and a delay to plug formation, except when the tube stiffness was very low, in which case the impact of including surfactant was minimal. They argued that this slowing implies an increase in the critical thickness required for plug formation, since simulations were run to a fixed finite time. This observation has relevance to mucus plug formation in airways, which typically form within the time scale of a single breath cycle. Experimental results have confirmed the decreased growth rates and increased times for plug formation to occur due to surfactant \citep{cassidy_1999_surfactant}. Computational fluid dynamics (CFD) simulations have also been conducted which, unlike quasi-one-dimensional models, can capture the post-coalescence phase of plug formation as well as the pre-coalescence phase \citep{romano_2022_surfactant}. It was found, again, that introducing surfactant delayed plug formation, and also that it decreased the stress on the tube wall during plugging. \textcolor{black}{The shear stress exerted on the tube wall is a physiologically important variable in airway closure models as it has been shown that epithelial cell damage can be caused by sufficiently large stresses being exerted on the airway wall \citep{huh_acoustically_2007}.}

There has been some attention on the effect of non-Newtonian liquid rheology on this flow in the case that there is no surfactant present. \citet{halpern_effect_2010} studied the effect of viscoelasticity, showing that the time for a plug to form can be decreased by increasing the Weissenberg number if the layer is sufficiently thick. \citet{romano_effect_2021} also investigated a viscoelastic version of the flow, using CFD, and found that elastic effects can induce a significant peak in the shear stress on the tube wall in the post-coalescence phase of plug formation. \citet{erken_2022_elastoviscoplastic} took a similar approach but with an elastoviscoplastic model for the liquid layer, showing that elastic effects can also impact how the fluid yields, particularly around and immediately after plug formation. 

\citet{shemilt_2022_surface} derived reduced-order models to study the effect of viscoplastic rheology on the dynamics of thin films and of thick films in the lead-up to plug formation. It was found that increasing the capillary Bingham number, $B$ (a parameter proportional to the liquid yield stress), can suppress instability by rigidification of the layer, reduce deformation when there is instability and significantly increase the critical layer thickness required for plug formation to occur. A model describing the evolution of thicker layers was developed using a long-wave approximation, and a simplified evolution equation was deduced using thin-film theory. 
The viscoplastic long-wave theory used is closely related to the planar thin-film theory exposed by \citet{BALMFORTH199965}. The structure of the flow is qualitatively the same in long-wave and thin-film theories: where the shear stress exceeds the yield stress the fluid is fully yielded and the flow is shear-dominated, but where the shear stress is less than the yield stress the flow is plug-like with no shear flow at leading order. However, due to changes in the surrounding flow, some regions with plug-like flow must still deform. In these regions, the yield stress is exceeded by an asymptotically small amount \citep{BALMFORTH199965} and the regions are referred to as `pseudo-plugs' \citep{walton_axial_1991}. Whilst the long-wave theory is similar to thin-film theory, it differs in a few crucial ways: the layer is not assumed to be thin relative to the tube radius, additional terms are retained in the evolution equations that more accurately capture the curvature of the geometry, and the exact expression for the curvature of the interface is used instead of the linearised version used in the thin-film theory. Thus, long-wave theory provides a composite approximation to the dynamics of a thick film, which is accurate where the layer is thin and also describes well the regions of the flow that are approximately capillary static. Equivalent, or very similar, approximations have been used previously to study this flow when the liquid layer is viscoelastic \citep{halpern_effect_2010} or Newtonian \citep{gauglitz_extended_1988,johnson_nonlinear_1991,camassa_2014_gravity,camassa_2017_air}, including when insoluble surfactant is present \citep{otis_effect_1990,otis_role_1993,ogrosky_linear_2021}. 

Viscoplastic thin-film theory has been used to describe various other interfacial flows where surface tension plays a key role. \citet{balmforth_surface_2007} studied the surface-tension-driven fingering instability in flow down an inclined plane, showing that the yield stress has a stabilising effect. \citet{jalaal_long_2016} developed a model using thin-film theory for a propagating bubble through a tube filled with viscoplastic fluid, which they validated against CFD simulations. Increasing the Bingham number was found to increase the thickness of the film between the bubble and the tube wall. Thin-film theory compared well with simulations for low Bingham numbers but less well when the film thickness increased. Thin-film models for the axisymmetric spreading of droplets \citep{jalaal_stoeber_balmforth_2021} and the spreading of long extruded filaments \citep{van_der_kolk_tieman_jalaal_2023} have been developed and used to predict the distances reached by spreading fronts, which decrease as the capillary Bingham number is increased. 
Whilst these studies all addressed the effect of viscoplastic rheology on capillary phenomena, surfactant effects were not incorporated in any of the models. \citet{craster_surfactant_2000} used thin-film theory to model surfactant-driven flow in viscoplastic films, demonstrating that after Marangoni forces cause a spreading front to develop, the yield stress can rigidify the layer before it returns to a uniform height profile. This study did not, however, include the effects of capillary forces. To the authors' knowledge, viscoplastic thin-film theory has not previously been used to study any flows where both capillary and Marangoni forces are present. 

In this study, we develop a model for the evolution of a liquid film coating the interior of a tube, where the flow is driven by surface tension but is also influenced by Marangoni forces. Our aim is to quantify the effects of surfactant on the capillary instability and on the previously established effects of the liquid's yield stress \citep{shemilt_2022_surface}. We will derive evolution equations for the layer height and surfactant concentration using long-wave theory, and subsequently deduce a simpler version of these equations that is valid in the thin-film limit. In order to focus attention on the interaction between Marangoni, capillary and yield-stress effects, other phenomena relevant to airway modelling are not included in the model. The Bingham constitutive model is used, which does not include rheological properties such as shear-thinning, viscoelasticity or thixotropy, but allows us to focus attention on yield-stress effects. Surface tension is assumed to vary linearly with surfactant concentration. For pulmonary surfactants this relation is nonlinear \citep{SCHURCH2001195}, but the linear model captures key Marangoni effects while being amenable to detailed analysis. This choice is in keeping with previous models \citep{halpern_surfactant_1993,ogrosky_linear_2021}. Moreover, gravity is assumed negligible, the tube wall is assumed rigid and the air in the centre of the tube is assumed passive. Numerical solutions of both the long-wave and thin-film equations will be used to elucidate the new features of the dynamic evolution that arise due to the presence of surfactant, and also to systematically explore parameter space. We will exploit the relative simplicity of the thin-film equations to study the behaviour of the layer in a late-time limit and in the limit of large Marangoni number (when surfactant is strong). By computing and analysing solutions of the long-wave equations, we will quantify how surfactant alters the dynamics leading to plug formation and the critical thickness required for plugging to occur. In particular, we will show how surfactant acts synergistically with the yield stress to stabilise the liquid layer. 

The paper will proceed as follows. In \S\ref{sec:formulation}, we derive evolution equations for the layer thickness and surfactant concentration, with the long-wave equations given in \S\ref{sec:LWmethods} and the thin-film equations in \S\ref{sec:TFmethods}. Solution methods will then be briefly discussed in \S\ref{sec:solnmethods}. Results from the thin-film system will be presented in \S\ref{sec:TF}. An example numerical simulation will be discussed in \S\ref{sec:TFexample}, late-time asymptotic analysis of the thin-film equations will be presented in \S\ref{sec:latetimeasymptotics} and the effect of varying the Marangoni number on the stability and evolution of the layer will be systematically addressed in \S\ref{sec:TFstability}. Results for thick films from the long-wave theory will be given in \S\ref{sec:LW}. An example numerical simulation will be examined in \S\ref{sec:LWexamples}, a discussion of the behaviour when the Marangoni number is large will be given in \S\ref{sec:LWlargeM} and the effect of surfactant on the critical thickness required for plug formation will be explored in \S\ref{sec:LWcritical}. Finally, in \S\ref{sec:discussion}, there will be a discussion of the significance of the results, particularly in relation to modelling airway closure in the lungs.

\section{\label{sec:formulation}Model formulation}

\subsection{\label{sec:governingeqns}Governing equations and boundary conditions}

\begin{figure*}
    \centering
    \includegraphics[width=\textwidth]{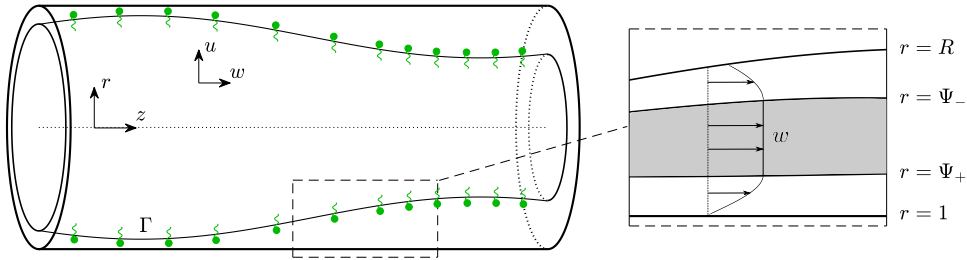}
    \caption{Left: sketch of the model geometry. The air-liquid interface is located at $r=R(z,t)$. Insoluble surfactant is present at the interface, with (non-dimensionalised) concentration $\Gamma$. Right: an illustration of a possible axial velocity profile in the liquid layer, ${w}$. Fully yielded, shear-dominated regions (white) are shown adjacent to the interface ($R\leq r\leq\Psi_-$) and adjacent to the wall ($\Psi_+\leq r\leq 1$), with a plug-like region (grey) in between ($\Psi_-<r<\Psi_+$). 
    }
    \label{fig:geometry}
\end{figure*}

We consider a rigid circular cylindrical tube lined on its interior by a layer of viscoplastic liquid, with a gas in the centre of the tube and insoluble surfactant present at the gas-liquid interface (see figure \ref{fig:geometry}). We assume the system is axisymmetric so it can be described using cylindrical coordinates $(r^*,z^*)$. The tube has radius $a$ and the interface is located at $r^* = R^*(z^*,t^*)$. The liquid layer has velocity $\boldsymbol{u}^*(r^*,z^*,t^*) = u^*\boldsymbol{\hat{r}}+w^*\boldsymbol{\hat{z}}$, and pressure $p^*(r^*,z^*,t^*)$ relative to the gas pressure, which is assumed to be spatially uniform. 

The liquid is incompressible and we assume inertia is negligible, so conservation of mass and momentum imply
\refstepcounter{equation}
$$
    \nabla^*\cdot\boldsymbol{u}^*=\boldsymbol{0},\quad\nabla^*\cdot\boldsymbol{\tau}^* = \nabla^* p^* \quad\mbox{for }\quad R^*\leq r^* \leq a,
    \eqno{(\theequation{\mathit{a},{b}})}\label{massmomStk}
$$
where $\boldsymbol{\tau}^*$ is the deviatoric stress tensor. The boundary conditions at the interface are the kinematic condition,
\begin{equation}
    \partial_t^*R^* + w^*\partial^*_zR^* = u^* \quad\mathrm{at}\quad r^*=R^*,
    \label{kinBCdim}
\end{equation}
and the stress condition,
\begin{equation}
    -p^*\boldsymbol{n} + \boldsymbol{\tau}^*\cdot\boldsymbol{n} = \sigma^*\kappa^*\boldsymbol{n} + \nabla_s^*\sigma^*,
    \label{normalBCsdim}
\end{equation}
where $\sigma^*$ is the surface tension, $\kappa^*$ is the curvature of the interface, $\boldsymbol{n}$ is the unit normal to the interface directed away from the liquid layer and $\nabla_s^*$ is the surface gradient operator. At the tube wall, there is no slip and no penetration,
\begin{equation}
    u^* = w^* = 0 \quad\mathrm{at}\quad r^* = a.
    \label{wallBCsdim}
\end{equation}
The liquid is assumed to be a Bingham fluid with constitutive relation,
\begin{equation}
\begin{array}{ll}
    \tau^*_{ij} = \left(\eta+\frac{\tau_y}{\dot\gamma^*}\right){\dot\gamma}^*_{ij} \quad \mathrm{if} \quad \tau^* \geq \tau_y,\\[8pt]
    {\dot\gamma}^*_{ij} = 0 \quad \mathrm{if}\quad \tau^*<\tau_y,
    \end{array}
    \label{constiteqndim}
\end{equation}
where $\eta$ is a viscosity, $\tau_y$ is the yield stress, $\boldsymbol{\dot\gamma}^* = \nabla^*\boldsymbol{u}^* + \nabla^*{\boldsymbol{u}^*}^T$ is the shear-rate tensor, and $\tau^*$ and $\dot\gamma^*$ are the second invariants of $\boldsymbol{\tau}^*$ and $\boldsymbol{\dot\gamma}^*$, respectively. The second invariant of a tensor $\boldsymbol{\mathsf{T}}$ is defined as $\mathsf{T} = \sqrt{\frac{1}{2}\mathsf{T}_{ij}\mathsf{T}_{ij}}$.


We take the surface tension of the interface to be a linear function of the surfactant concentration,
\begin{equation}
    \sigma^* = \sigma_0 + K(\Gamma_0-\Gamma^*),
    \label{sigmastar}
\end{equation}
where $\sigma_0$ and $\Gamma_0$ represent the constant values of the surface tension and surfactant concentration in an unperturbed state, and $K>0$ is a constant. We assume that the surfactant is insoluble so its motion is governed by the transport equation \citep[see, e.g.,][]{stone_1990_derivation}
\begin{equation}
    \partial^*_t\Gamma^* + \nabla_s^*\cdot(\Gamma^*\boldsymbol{u}^*_s)  + \Gamma^*\kappa^*(\boldsymbol{u}^*\cdot\boldsymbol{n})= 0,
    \label{TransportEqnDim}
\end{equation}
where $\boldsymbol{u}_s^*$ is the velocity along the interface. In \eqref{TransportEqnDim}, we have neglected diffusion of surfactant in order to focus on the limit of advection-dominated transport. In lung airways, although there is significant variation in measurements of the properties of surfactant and mucus, we typically expect surfactant diffusivity to be small and that advection will be the dominant transport mechanism \citep{craster_surfactant_2000,lai_micro-_2009,chen_determination_2019}. At the lateral boundaries of the domain, we impose symmetry boundary conditions, 
\begin{equation}
    \partial^*_zR^* = \tau^*_{rz} = w^*=(\boldsymbol{u}^*_s\cdot\boldsymbol{\hat{z}})\Gamma = 0
    \quad\mathrm{at}\quad z^*=\{0,L^*\},
    \label{sideBCs1}
\end{equation}
which enforce that the first derivative of the layer height is zero and that there is no flux of fluid or surfactant across the side boundaries. 

\subsection{Non-dimensionalisation}

To non-dimensionalise \eqref{massmomStk}-\eqref{TransportEqnDim}, we introduce the dimensionless quantities,
\begin{equation}
    \left.\begin{array}{c}
    \displaystyle
        (r,z) = \left(\frac{r^*}{a},\frac{z^*}{a}\right), \quad{t} = \frac{\sigma_0}{a\eta}t^*, \quad \boldsymbol{\dot\gamma} = \frac{a\eta}{\sigma_0}\boldsymbol{\dot\gamma}^*, \quad
        ({u},{w}) = \frac{\eta}{\sigma_0}(u^*,w^*),\quad \Gamma = \frac{\Gamma^*}{\Gamma_0}, \\[14pt]
        \displaystyle
        \boldsymbol{\tau} = \frac{a}{\sigma_0}\boldsymbol{\tau}^*,\quad
        R = \frac{R^*}{a}, \quad \sigma = \frac{\sigma^*}{\sigma_0}, \quad  p = \frac{a}{\sigma_0}p^*, \quad\kappa = a\kappa^*,\quad L = \frac{L^*}{a}.
        \end{array}\right\}
        \label{nondim}
\end{equation}
The mass and momentum conservation equations\ \eqref{massmomStk} become
\begin{eqnarray}
    0 &=& \partial_z{w} + \frac{1}{r}\partial_r(r{u}),\label{masscons}\\
    \partial_r{p} &=& \frac{1}{r}\partial_r(r\tau_{rr})+\partial_z\tau_{rz}-\frac{\tau_{\theta\theta}}{r},\\
    \partial_z{p} &=& \frac{1}{r}\partial_r(r\tau_{rz}) + \partial_z\tau_{zz}.\label{horizmom}
\end{eqnarray}
The wall boundary conditions \eqref{wallBCsdim} are
\begin{equation}
    {u} = {w} = 0 \quad \mbox{on} \quad r = 1,
\end{equation}
the kinematic boundary condition \eqref{kinBCdim} is
\begin{equation}
    \partial_{{t}}R+{w}\partial_zR = {u} \quad\mbox{on}\quad r=R,
    \label{kinBCnodim}
\end{equation}
and the stress boundary condition \eqref{normalBCsdim} is
\begin{equation}
    -{p}n_i + \tau_{ij}n_j = \sigma\kappa n_i + (\delta_{ij}-n_in_j)\partial_j\sigma \quad\mbox{on}\quad r=R,
\end{equation}
where the curvature of the interface is
\begin{equation}
    \kappa = \frac{1}{\sqrt{1+(\partial_zR)^2}}\left[\frac{1}{R}-\frac{\partial_{zz}R}{1+(\partial_zR)^2}\right].
    \label{kappa}
\end{equation}
The constitutive relation \eqref{constiteqndim} is
\begin{equation}
\begin{array}{ll}
    \tau_{ij} = \left(1+\frac{\B}{\dot\gamma}\right){\dot\gamma}_{ij} \quad \mathrm{if} \quad \tau \geq \B,\\[8pt]
    {\dot\gamma}_{ij} = 0 \quad \mathrm{if}\quad \tau<\B,
    \end{array}
    \label{constitnodim}
\end{equation}
where
\begin{equation}
    {\B}=\frac{a\tau_y}{\sigma_0}
\end{equation} 
is a capillary Bingham (or plastocapillarity) number \citep{jalaal_stoeber_balmforth_2021,van_der_kolk_tieman_jalaal_2023}. The equation of state \eqref{sigmastar} becomes
\begin{equation}
    \sigma(\Gamma) = 1 + {\Ma}(1-\Gamma),\label{eqnofstate}
\end{equation}
where 
\begin{equation}
\Ma = \frac{K\Gamma_0}{\sigma_0}
\end{equation} 
is a Marangoni number. The transport equation \eqref{TransportEqnDim} can be expressed in conservative form \citep{halpern_frenkel_2003} as 
\begin{equation}
\partial_{{t}}\left[R\Gamma\sqrt{1+(\partial_zR)^2}\right] + \partial_z\left[{w}_sR\Gamma\sqrt{
1+(\partial_zR)^2}\right] = 0,
\label{surftransport1}
\end{equation}
where $w_s$ is the axial component of the surface velocity. We now examine simplified versions of \eqref{masscons}-\eqref{surftransport1}, first in a long-wave limit, appropriate for thicker films, and then in a more restrictive thin-film limit. 

\subsection{Long-wave theory\label{sec:LWmethods}}

We consider the system \eqref{masscons}-\eqref{surftransport1} in a long-wave limit by defining a characteristic axial length scale, $\mathcal{L}=a/\delta$, where $\delta\ll1$. We define stretched variables, 
\begin{equation}
    \Bar{z}=\delta z,\quad \Bar{u}=\frac{u}{\delta^2},\quad \Bar{w}=\frac{w}{\delta},\quad \Bar{t}=\delta^2 {t},\quad \bar{\boldsymbol{\tau}} = \frac{\boldsymbol{\tau}}{\delta},\quad \bar{L} = \delta L,
    \label{LWscaling}
\end{equation}
\textcolor{black}{where the choice of scalings for $z$ and $L$ arise from the long-wave approximation, the scaling for $\boldsymbol{\tau}$ is then chosen so that the radial gradient of shear stress balances the axial pressure gradient in \eqref{horizmom}, the scaling for $w$ is chosen so that the shear stress balances with the corresponding term in the strain-rate tensor in \eqref{constitnodim}, $u$ is scaled such that the mass conservation equation \eqref{masscons} balances and, finally, the scaling for $t$ allows all terms in the kinematic boundary condition \eqref{kinBCnodim} to balance.} (The scalings \eqref{LWscaling} correct those printed in equation (2.11) in \citet{shemilt_2022_surface} but the equations derived there are still correct and consistent with what we derive here.) We then truncate the governing equations \eqref{masscons}-\eqref{surftransport1} at leading order in $\delta$, with the exception of \eqref{kappa} where we retain the exact curvature, which is a commonly used device to improve accuracy in near-static regions of the flow \citep{gauglitz_extended_1988,halpern_fluid-elastic_1992,halpern_surfactant_1993,halpern_effect_2010,ogrosky_linear_2021,shemilt_2022_surface}. The leading-order governing equations and boundary conditions are given in Appendix \ref{app:LWderivation}, where we also detail the derivation of the long-wave evolution equations, which are presented in the remainder of this section. As has been done previously in similar problems \citep{camassa_ring_2012,shemilt_2022_surface,camassa_viscous_2015,ogrosky_linear_2021}, we will present the long-wave equations here in terms of the unscaled variables defined in \eqref{nondim} instead of the scaled variables \eqref{LWscaling}, but the equations still represent the leading-order theory in the limit $\delta\ll1$.
 
 To understand the structure of the evolution equations, it is important to first understand the structure of the flow and the ways in which the layer can yield. Where the magnitude of the shear stress is larger than the yield stress, $|\tau_{rz}|>\B$ for some $R(z,t)<r<1$, the fluid is fully yielded and the flow is shear-dominated. Where $|\tau_{rz}|\leq \B$ for some $R(z,t)<r<1$, the flow is plug-like and the leading-order axial velocity is independent of $r$, i.e. ${w}={w}_p(z,t)$. If ${w}_p$ is also independent of $z$ in a plug-like region, then it is a rigid plug. If not, then that plug-like region is deforming axially so it must be yielded. In those yielded plug-like regions, the normal stresses become leading-order in $\delta$ and combine with the shear stress such that the yield condition is just met; such a region is referred to as `pseudo-plug' \citep{walton_axial_1991}. This general structure is the same as in viscoplastic thin-film flows, as detailed by \citep{BALMFORTH199965}. For the remainder of this section, subscripts will be used to denote derivatives. 
 
\begin{figure*}
    \centering
    \includegraphics[width=\textwidth]{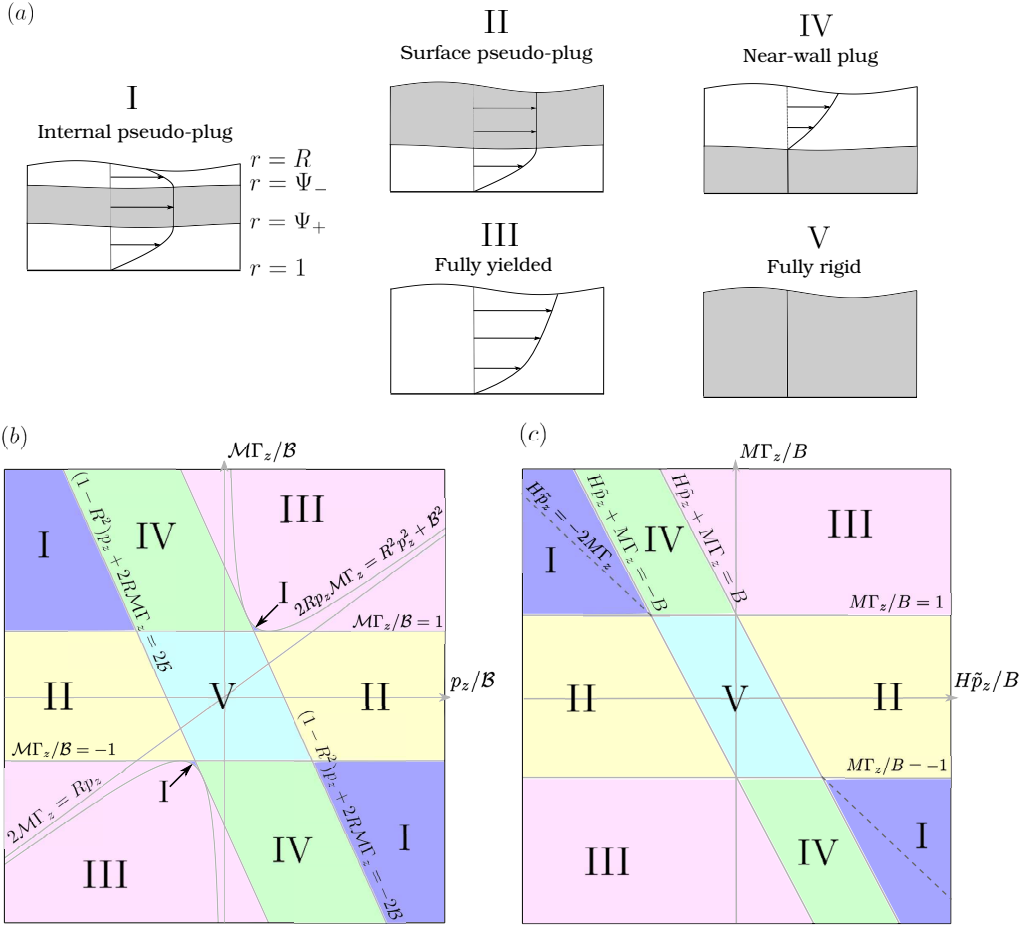}
    \caption{(a) The fives types of yielding that can occur in the layer. Plug-like regions are shown in grey and fully-yielded regions in white. Typical axial velocity profiles are also sketched. Below are maps of parameter space showing where these yielding types occur in (b) the long-wave system and (c) the thin-film system. When plotting the map in (b), we treat $R$ as a fixed parameter in order to focus on variation with $p_z$ and $\Ma\Gamma_z$. In (b), as in \S \ref{sec:LWmethods}, the parameter map is plotted in terms of the unscaled variables \eqref{nondim}. In (c), the map is plotted in terms of the scaled thin-film variables introduced in \eqref{TFscaling}. Along the dashed line in (c), the surface velocity is exactly zero, $\tilde{w}_s=0$.}
    \label{fig:yieldmap}
\end{figure*}

 Where the fully-yielded regions, pseudo-plugs and rigid plugs occur in the liquid film depends on the competition between surface and bulk forcing, via $\Ma\Gamma_z$ and $p_z$, and their sizes relative to $\B$. \color{black} In Appendix \ref{app:LWderivation}, we show that, in the long-wave limit, the capillary pressure is given by
 \begin{equation}
     p = -\kappa\left[1+\Ma(1-\Gamma)\right],
     \label{pLW}
 \end{equation}
 where the curvature of the interface, $\kappa$, is defined in \eqref{kappa}, and that the shear stress is given by 
 \begin{equation}
 \tau_{rz} = \frac{p_z}{2}\left(r - \frac{R^2}{r}\right) + \frac{R}{r}\mathcal{M}\Gamma_z.\label{taurzmain}
 \end{equation}
By noting how $\tau_{rz}$ depends on $r$ in \eqref{taurzmain}, it can be deduced that there are five qualitatively different types of yielding that can occur in the layer, corresponding to the five possible combinations of fully-yielded regions, rigid plugs and pseudo-plugs that can exist in $R\leq r\leq1$. 
  We define two internal surfaces, $r=\Psi_-(z,t)$ and $r=\Psi_+(z,t)$, which separate fully-yielded and plug-like regions. The fully-yielded regions (where $|\tau_{rz}|>\mathcal{B}$) occupy $R\leq r\leq\Psi_-$ and $\Psi_+\leq r\leq1$, and the plug-like region (where $|\tau_{rz}|\leq\mathcal{B}$) occupies $\Psi_-\leq r\leq\Psi_+$.  \color{black} Note that each of these three regions may not always exist, in which case the width of the region will be zero. The five possible types of yielding are as follows and are illustrated in figure \ref{fig:yieldmap}(a).
 \vspace{4pt}
\begin{enumerate}[label=\Roman*.]
    \item\, Internal pseudo-plug ($R<\Psi_-<\Psi_+<1$). Here, fully-yielded regions adjacent to the wall and the interface are separated by a pseudo-plug.\\
    \item\, Surface pseudo-plug ($\Psi_-= R<\Psi_+<1$). Here, the pseudo-plug extends to the interface and the only fully-yielded region is adjacent to the wall.\\
    \item\, Fully yielded ($\Psi_-=\Psi_+= R$ or $1=\Psi_-=\Psi_+$). In this case, there is no plug-like region and the whole layer is fully yielded.\\
    \item\, Near-wall plug ($R<\Psi_-<1=\Psi_+$). Here, there is a rigid plug adjacent to the wall and the only fully-yielded region is adjacent to the interface.\\
    \item\, Fully rigid ($\Psi_-= R$ and $\Psi_+=1$). Neither yielded region exists and the whole layer is a rigid plug.\\
\end{enumerate}
\vspace{4pt}
\citet{hewitt_viscoplastic_2012} identified these same five yielding types in the flow of a thin film between two moving solid surfaces. In that problem, there are simple criteria to determine which yielding type occurs based on the size of the shear stress at the two solid boundaries. Here, the additional complexity of the long-wave theory compared to the thin-film theory means there are not such simple criteria. Instead, we derive the following expressions for $\Psi_\pm$, from which the type of yielding can be deduced. We find
\begin{equation}
    \Psi_\pm = \max\left[R,\min\left(1,\psi_\pm\right)\right],\label{Psipmdefn}
\end{equation}
where, if $p_z\neq0$, we have
    \begin{equation}
        \psi_{\pm} = \left\{\begin{array}{ll}
        \pm\frac{\B}{|p_z|}+\sqrt{\left(\frac{\B}{p_z}\right)^2+R^2-\frac{2R\Ma\Gamma_z}{p_z}}&\mbox{if}\quad \frac{2\Ma\Gamma_z}{Rp_z}<1,\vspace{4pt}\\
        \vspace{4pt}
        \frac{\B}{|p_z|}\pm\sqrt{\left(\frac{\B}{p_z}\right)^2+R^2-\frac{2R\Ma\Gamma_z}{p_z}} & \mbox{if}\quad 1+\frac{\B^2}{R^2p_z^2}\geq\frac{2\Ma\Gamma_z}{Rp_z}\geq1,\\
        R& \mbox{if}\quad \frac{2\Ma\Gamma_z}{Rp_z}>1+\frac{\B^2}{R^2p_z^2},
        \end{array}\right.
        \label{psipmLW}
    \end{equation}
and, if $p_z=0$, then $\psi_-=R\Ma|\Gamma_z|/\B$ and $\psi_+=1$. Note that these definitions mean that $\Psi_\pm$ are continuous everywhere, including at $p_z=0$. 

Given values of $R$, $\B$, $p_z$ and $\Ma\Gamma_z$, the type of yielding can then be deduced from \eqref{psipmLW}. A more intuitive illustration of when the yielding types I-V occur is given by figure \ref{fig:yieldmap}(b). It shows how the yielding type depends on the size of the capillary stress relative to the yield stress, via $p_z/\B$, and on the size of the Marangoni stress relative to the yield stress, via $\Ma\Gamma_z/\B$. We plot figure \ref{fig:yieldmap}(b) assuming $R$ is fixed, even though in general it will vary with $z$. We now briefly survey $(p_z/\B,\Ma\Gamma_z/\B)$-space. Starting in the upper left region of figure \ref{fig:yieldmap}(b), where capillary and Marangoni stresses are both large but with opposite signs, there is yielding of type I. Suppose $\Ma\Gamma_z/\B$ is then decreased, so that we cross the line $\Ma\Gamma_z/\B=1$. Then the shear stress at the interface has dropped below $\B$ so there can no longer be a fully-yielded region at the interface, and the layer then exhibits yielding of type II. Decreasing $\Ma\Gamma_z/\B$ further so that $\Ma\Gamma_z/\B<-1$, the Marangoni stress at the interface again exceeds the yield stress so there must be yielding at the interface, but it now acts in the same direction as the capillary stress. The layer then exhibits yielding of type III. Now increasing $p_z/\B$, we remain in yielding type III until we reach the line $(1-R^2)p_z+2R\Ma\Gamma_z=-2\B$. After crossing this line, the capillary stress is no longer strong enough to yield the fluid adjacent to the wall, so a rigid plug develops there and we have yielding of type IV. Increasing $p_z/\B$ yet further, so that we cross the line $(1-R^2)p_z+2R\Ma\Gamma_z=2\B$, now $p_z/\B$ is large enough that the lower fully-yielded region appears again so we have returned to yielding of type I, but with the flow in the opposite direction to when we started. Symmetry means that if we proceed in the same fashion around the other half of the plane, the yielding transitions will be the same as just described. There are two regions in figure \ref{fig:yieldmap} that we have not yet discussed. When $p_z/\B$ and $\Ma\Gamma_z/\B$ are both small then there is no yielding, so there is a region with yielding of type V around the origin. Finally, yielding of type I can also be observed in two small regions near $(p_z/\B,\Ma\Gamma_z/\B)=(\pm2/(1-R),\pm1)$. The existence of these relies on the shear stress in the long-wave theory being nonlinear, so they do not exist in the simpler thin-film limit (figure \ref{fig:yieldmap}c). 

Given the expressions for $\Psi_\pm$ in \eqref{Psipmdefn} and \eqref{psipmLW}, we can now present the long-wave evolution equations, which are derived in Appendix \ref{app:LWderivation}. The evolution equation for the interface position, $R$, is
\begin{equation}
    R_t=\frac{1}{R}Q_z,\label{LWevoleqn}
\end{equation}
where the axial volume flux is given by
    \begin{equation}
        Q = \left\{\begin{array}{ll}
        -\frac{p_z}{16}F_1-\frac{1}{4}R\Ma\Gamma_zF_2-\frac{\B}{6}\sgn(p_z)(F_3+F_4)&\mbox{if}\quad \frac{2\Ma\Gamma_z}{Rp_z}<1,\vspace{4pt}\\
        \vspace{4pt}
        -\frac{p_z}{16}F_1-\frac{1}{4}R\Ma\Gamma_zF_2-\frac{\B}{6}\sgn(p_z)(F_3-F_4) & \mbox{if}\quad \frac{2\Ma\Gamma_z}{Rp_z}\geq 1,\\
        -\frac{1}{4}R\Ma\Gamma_zF_2 + \frac{\B}{6}\sgn(\Gamma_z)F_4& \mbox{if}\quad p_z=0,
        \end{array}\right.
        \label{Q}
    \end{equation}
with
\begin{subeqnarray}
F_1 &=& \Psi_-^4-4R^2\Psi_-^2+R^4\left[3-4\log\left(\frac{R\Psi_+}{\Psi_-}\right)\right]
        -\Psi_+^4 + 4R^2\Psi_+^2-4R^2+1,\quad\quad\quad
        \label{F1}\\[3pt]
        F_2 &=& \Psi_-^2-R^2-\Psi_+^2+1+2R^2\log\left(\frac{R\Psi_+}{\Psi_-}\right),
        \label{F2}\\[3pt]
        F_3 &=& (\Psi_+-1)\left(-3R^2+1+\Psi_++\Psi_+^2\right),\quad F_4=\Psi_-^3-3R^2\Psi_-+2R^3.
        \label{F3F4}
\end{subeqnarray}
If $\Ma=0$, or equivalently if there is no surfactant, $\Gamma=0$, then $\Psi_-=R$ and \eqref{LWevoleqn}-\eqref{F3F4} reduces to the evolution equation for the surfactant-free problem (correcting a typographical sign error in the expression for $Q$ given in equation (2.16) in \cite{shemilt_2022_surface}). The surfactant transport equation is
\begin{equation}
    (R\Gamma)_t + (w_sR\Gamma)_z=0,\label{surftransportLW}
\end{equation}
where the surface velocity is
    \begin{equation}
        w_s = \left\{\begin{array}{ll}
        \frac{1}{4}p_zG_1+R\Ma\Gamma_zG_2+\B\sgn(p_z)(G_3+G_4)&\mbox{if} \quad \frac{2\Ma\Gamma_z}{Rp_z}<1,\vspace{4pt}\\
        \vspace{4pt}
        \frac{1}{4}p_zG_1+R\Ma\Gamma_zG_2+\B\sgn(p_z)(G_3-G_4) & \mbox{if}\quad \frac{2\Ma\Gamma_z}{Rp_z}\geq1,\\
        R\Ma\Gamma_zG_2 - \B\sgn(\Gamma_z)G_4& \mbox{if}\quad p_z=0,
        \end{array}\right.
        \label{wsLW}
    \end{equation}
with
\refstepcounter{equation}
    $$
        G_1 = R^2+\Psi_+^2-\Psi_-^2-1-2R^2\log\left(\frac{R\Psi_+}{\Psi_-}\right),
        \eqno{(\theequation{\mathit{a}})}
        \label{G1}
    $$
    $$
        G_2 = \log\left(\frac{R\Psi_+}{\Psi_-}\right),\quad G_3 = 1-\Psi_+,\quad G_4 = R-\Psi_-.
        \eqno{(\theequation{\mathit{b,c,d}})}
        \label{G2G3G4}
    $$
Equations \eqref{LWevoleqn}-\eqref{G2G3G4} are solved subject to the boundary conditions
\begin{equation}
    R_z = Q = w_s\Gamma = 0 \quad\mbox{at}\quad z=\{0,L\},
    \label{LWsideBCs}
\end{equation}
which completes the long-wave system of equations. Finally, by evaluating the shear stress \eqref{Ataurz} at $r=1$, we can deduce an expression for the stress exerted on the tube wall,
\begin{equation}
    \tau_w = \frac{p_z}{2}(1-R^2) + R\Ma\Gamma_z.
    \label{LWwallstress}
\end{equation}


\subsection{Thin-film theory\label{sec:TFmethods}}

We now consider the system in a thin-film limit in order to derive simpler evolution equations that are more amenable to detailed analysis. In the thin-film theory, we assume that $|1-R|\ll1$, but no longer require that $\delta\ll1$. 
Rather than presenting a derivation of the thin-film equations from \eqref{masscons}-\eqref{surftransport1}, here we will show how they can be deduced directly from the long-wave equations \eqref{LWevoleqn}-\eqref{LWsideBCs}. The flow structure is qualitatively the same and the same five possible yielding types I-V also occur.  

We assume that
\begin{equation}
	R(z,t) = 1 - \epsilon H(z,t)
	\label{Hdefn}
\end{equation}
where $\epsilon\ll1$ and $H$ is the scaled layer thickness. Similarly, we define the boundaries between fully-yielded and plug-like regions,
\begin{equation}
	\Psi_\pm = 1 - \epsilon Y_\mp.
	\label{Ypmdefn}
\end{equation}
Since $\Psi_+\geq\Psi_-$, the definitions \eqref{Ypmdefn} mean $Y_+\geq Y_-$. Other relevant variables are rescaled by defining
\begin{equation}
	\Tilde{p} = \frac{1+p}{\epsilon},\quad\Tilde{\kappa} = \frac{\kappa-1}{\epsilon}\quad \Tilde{t}=\epsilon^3t, \quad \textcolor{black}{\tilde{\boldsymbol{\tau}} = \frac{\boldsymbol{\tau}}{\epsilon^2}},
	\label{TFscaling}
\end{equation}
and the scaled capillary Bingham and Marangoni numbers are given, respectively, by
\begin{equation}
	{B} = \frac{\B}{\epsilon^2} = \frac{a\tau_y}{\sigma_0\epsilon^2}\quad\mathrm{and}\quad
	M = \frac{\Ma}{\epsilon^2} = \frac{K\Gamma_0}{\sigma_0\epsilon^2}.
 \label{BMadefns}
\end{equation}
We then insert \eqref{Hdefn}-\eqref{BMadefns} into the long-wave equations \eqref{pLW}-\eqref{LWsideBCs} and truncate at leading order in $\epsilon$. From \eqref{pLW}, the thin-film capillary pressure gradient is given by
\begin{equation}
    \tilde{p}_z = -\tilde\kappa_z = -H_z-H_{zzz}
    \label{ptilde}
\end{equation}
to leading order in $\epsilon$. (Unlike in the thick-film case \eqref{pLW}, surfactant has no effect on the mean surface tension at leading order in the thin-film limit.) Note that with the scalings \eqref{TFscaling} and \eqref{BMadefns}, $2\Ma\Gamma_z/R{p}_z=O(\epsilon)$, so in the thin-film theory we always have $2\Ma\Gamma_z/R{p}_z<1$. This simplifies \eqref{psipmLW} somewhat, giving $Y_\pm=\max\left[0,\min\left(H,\mathcal{Y}_\pm\right)\right]$ where
\begin{equation}
    \mathcal{Y}_\pm = H + \frac{\textcolor{black}{M}\Gamma_z}{p_z}\pm\frac{\textcolor{black}{B}}{|p_z|},\label{Ydefns}
\end{equation}
assuming for now that $p_z\neq0$. As in \S \ref{sec:LWmethods}, we can present criteria for each of the five yielding types to occur based on the values of $Y_\pm$ (as in figure \ref{fig:yieldmap}a): I, internal pseudo-plug ($0<Y_-<Y_+<H$); II, surface pseudo-plug ($0<Y_-<H=Y_+$); III, fully yielded ($Y_\pm=0$ or $Y_\pm=H$); IV, near-wall plug ($Y_-=0<Y_+<H$); V, fully rigid ($Y_-=0$ and $Y_+=H$). Figure \ref{fig:yieldmap}(c) illustrates where in $(H\tilde{p}_z/B,M\Gamma_z/B)$-space each of types I-V occurs. The picture is similar to figure \ref{fig:yieldmap}(b), which was described in detail above, but is somewhat simplified. 

The evolution equation \eqref{LWevoleqn}, to leading order in $\epsilon$, becomes
\begin{equation}
	H_{\tilde{t}} + q_z = 0,
	\label{TFevoleqn}
\end{equation}
where the scaled axial volume flux is
\begin{multline}
    q = -\frac{1}{3}\tilde{p}_z\left[H^3+(H-Y_+)^3-(H-Y_-)^3\right]-\frac{1}{2}M\Gamma_z\left[H^2-(H-Y_-)^2+(H-Y_+)^2\right] \\+\frac{1}{2}B\sgn(\tilde{p}_z)\left[H^2-(H-Y_-)^2-(H-Y_+)^2\right],
    \label{qdefnTF}
\end{multline}
if $\tilde{p}_z\neq0$. When $\tilde{p}_z=0$, the flux is $q=-\frac{1}{2}\sgn(\Gamma_z)H^2(|M\Gamma_z|-B)$ if $|M\Gamma_z|>B$, and $q=0$ if $|M\Gamma_z|\leq B$. As can be seen in figure \ref{fig:yieldmap}(c), when $\tilde{p}_z=0$, the only type of yielding possible is type III, which occurs if $|M\Gamma_z|>B$; the layer is rigid (type V) if $|M\Gamma_z|\leq B$. 

The surfactant transport equation \eqref{surftransportLW}, at leading order, becomes
\begin{equation}
    \Gamma_{\tilde{t}} + \left[\tilde{w}_s\Gamma\right]_z = 0,\label{surftransportTF}
\end{equation}
where the scaled surface velocity is
\begin{equation}
    \tilde{w}_s = -\frac{1}{2}\tilde{p}_z\left[H^2+(H-Y_+)^2-(H-Y_-)^2\right] - M\Gamma_z(H+Y_--Y_+)-B\sgn(\tilde{p}_z)(H-Y_--Y_+),\label{wsTF}
\end{equation}
if $\tilde{p}_z\neq0$. When $\tilde{p}_z=0$, the surface velocity is $\tilde{w}_s = -\sgn(\Gamma_z)H(|M\Gamma_z|-B)$ if $|M\Gamma_z|>B$ and $\tilde{w}_s=0$ if $|M\Gamma_z|\leq B$. 
From \eqref{Ydefns} and \eqref{wsTF}, we can deduce that if $H\tilde{p}_z=-2M\Gamma_z$ then $Y_-=H-Y_+$ and $\tilde{w}_s=0$, meaning that the interface of the layer is immobilised. This line is included on figure \ref{fig:yieldmap}(c), and will form part of our later discussion. 
The lateral boundary conditions \eqref{LWsideBCs} reduce in the thin-film limit to
\begin{equation}
    H_z = q = \Gamma \tilde{w}_s = 0 \quad\mathrm{at} \quad z= \{0,L\}.
    \label{TFsideBCs}
\end{equation}
Equations \eqref{Ydefns}-\eqref{wsTF} with boundary conditions \eqref{TFsideBCs} comprise the thin-film system. The wall shear stress \eqref{LWwallstress}, in the thin-film limit, becomes
\begin{equation}
    \tilde{\tau}_w = H\tilde{p}_z + M\Gamma_z.
\end{equation}

\subsection{\label{sec:solnmethods}Solution methods}

When solving the thin-film or long-wave equations numerically, we use initial conditions with a uniform surfactant concentration across the layer and a perturbation to the interface height with wavelength $2L$. For the thin-film equations, these initial conditions are
\begin{equation}
    H(z,t=0) = 1 - A \cos\left(\frac{\pi z}{L}\right),\quad \Gamma(z,t=0) = 1,\label{TFICs}
\end{equation}
and for the long-wave equations, the equivalent conditions are
\begin{equation}
    R(z,t=0) = \sqrt{(1-\epsilon)^2-\epsilon^2 A^2/2}-\epsilon A\cos\left(\frac{\pi z}{L}\right), \quad \Gamma(z,t=0) = 1,\label{LWICs}
\end{equation}
where $A$ is the amplitude of the initial perturbation and $\epsilon$ is the ratio of the average layer thickness to the tube radius when $A=0$. The constant term in \eqref{LWICs} ensures that the volume of fluid is independent of $A$. As noted for the surfactant-free problem \citep{shemilt_2022_surface}, when the fluid is viscoplastic there is no linear instability since a finite-amplitude initial perturbation is required to yield. 

For ease of comparison with the surfactant-free problem \citep{shemilt_2022_surface} and other related studies \citep{gauglitz_extended_1988,halpern_effect_2010}, we solve the equations in a domain of length $\textcolor{black}{L=\sqrt{2}\pi}$. This is the domain length that can accommodate the most unstable mode in the Newtonian linear stability analysis \citep{hammond_nonlinear_1983}. The initial perturbation applied to the layer then corresponds to the one unstable Fourier mode that exists in the domain. Although this value of $L$ is not necessarily of unique importance in the viscoplastic problem, we do not observe any qualitative changes in the results when $L$ is changed by relatively small amounts.  
As in the surfactant-free problem, since there is nothing in the model equations that could break symmetry around the lateral boundaries of the domain, the boundary conditions \eqref{LWsideBCs} and \eqref{TFsideBCs} yield the same solutions as would be found using periodic boundary conditions. Symmetry rather than periodic boundary conditions allows a finer grid to be used for the spatial finite differencing at the same computational expense, since the computational domain is shorter.

To solve both the long-wave equations and the thin-film equations numerically, we use a regularisation introduced by \citet{jalaal_thesis_2016}, which was used for the surfactant-free problem \citep{shemilt_2022_surface} and has also been used for several other viscoplastic thin-film problems \citep{balmforth_surface_2007,jalaal_stoeber_balmforth_2021}. We define $\hat{Y}_\pm=\max\left(Y_{\min},Y_\pm\right)$ and $\hat{\Psi}_\pm=\min\left(1-Y_{\min},\Psi_\pm\right)$ where $Y_{\min}$ is a small parameter, and replace $Y_\pm$ and $\Psi_\pm$ with $\hat{Y}_\pm$ and $\hat{\Psi}_\pm$, respectively, in the equations. We choose $Y_{\min}=10^{-8}$ for all simulations, which is small enough that the exact value does not affect the results. \textcolor{black}{To solve the resulting equations, the domain $0\leq z\leq L$ is discretised into a grid of $N$ evenly spaced points (we typically use $N=200$), the spatial derivatives are approximated by second-order central finite differences and then the equations are time-stepped using an ODE solver in \textsc{Matlab}.}

\section{\label{sec:TF}Thin Films}

\subsection{Time evolution of a surfactant-laden thin film\label{sec:TFexample}}

\begin{figure*}

\includegraphics[width=\textwidth]{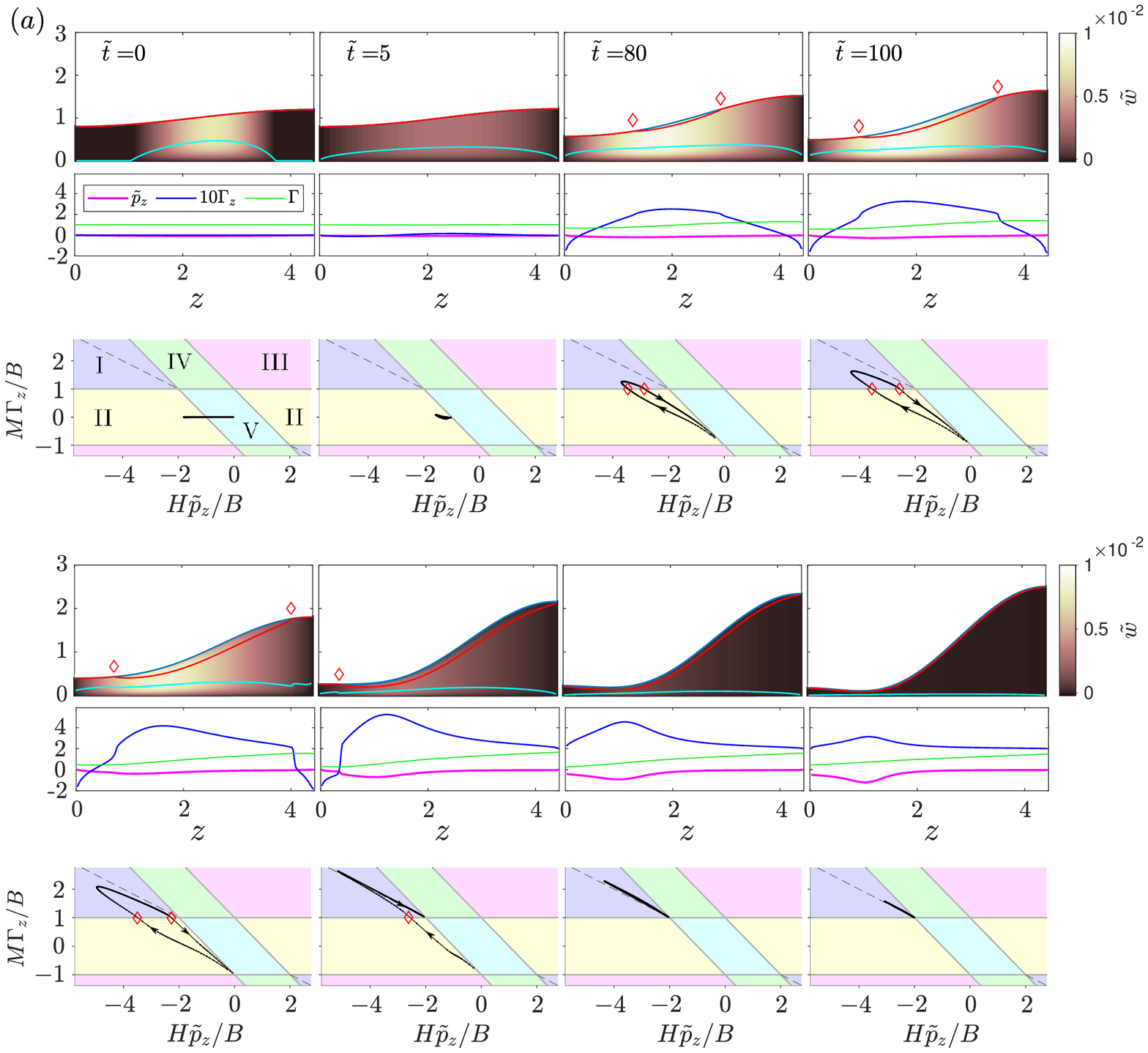}\vspace{10pt}

\includegraphics[width=0.95\textwidth]{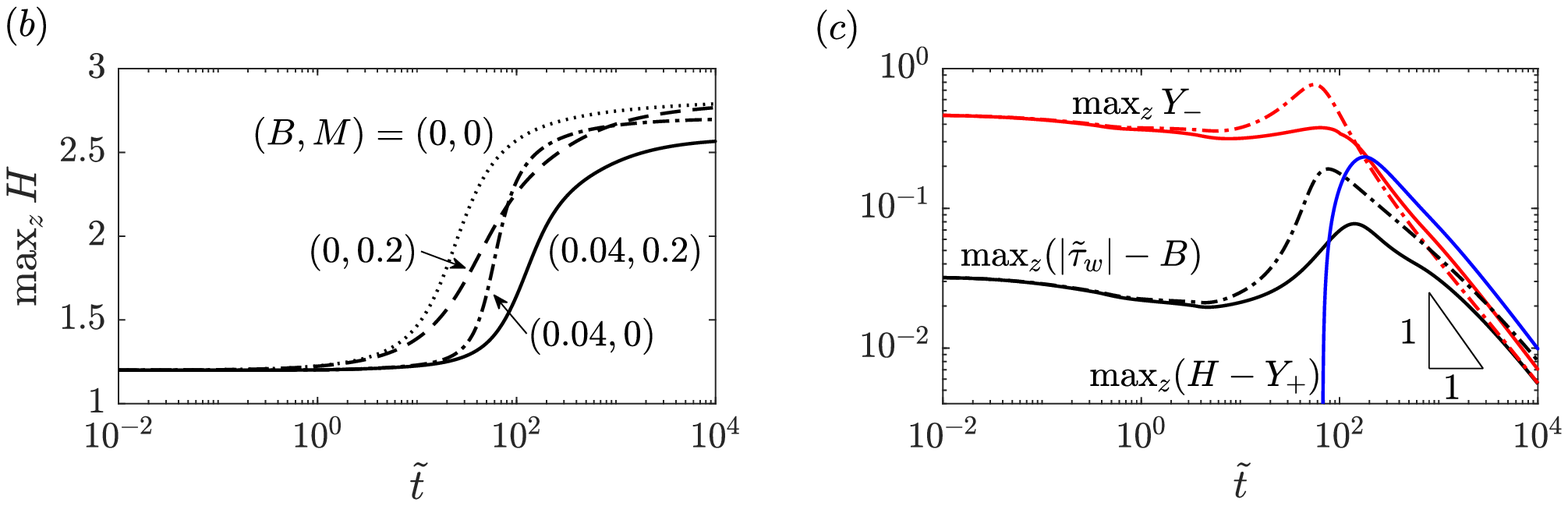}
\caption{\label{fig:TFevolution} (a) Snapshots from a numerical solution of the thin-film evolution equations \eqref{Ydefns}-\eqref{TFsideBCs} with $B=0.04$, $M=0.2$, $A=0.2$ at $\tilde{t}=\{0,5,80,100,130,250,500,2000\}$. At each $\tilde{t}$, there are three panels: the top panel shows the layer evolving, with $Y_-$ (cyan) and $Y_+$ (red), and the thin-film axial velocity $\tilde{w}$ represented by the colour map; the middle panel shows plots of $\tilde{p}_z$ (magenta), $\Gamma$ (green) and $10\Gamma_z$ (blue); the bottom panel shows the solution in $(H\tilde{p}_z/B$,$M\Gamma_z/B$)-space, in dotted black lines, with the dots corresponding to points evenly spaced along the domain $0<z<L$ and the arrows indicating the direction of increasing $z$. \textcolor{black}{Red diamonds on the first and third panels mark the boundaries of the region where there is yielding at the interface. (b) Time evolution of $\max_zH$ from the same simulation (solid), compared to the evolution of $\max_zH$ from a Newtonian surfactant-laden simulation with $(B,M)=(0,0.2)$ (dashed), a surfactant-free viscoplastic simulation with $(B,M)=(0.04,0)$ (dot-dashed) and a surfactant-free Newtonian simulation with $(B,M)=(0,0)$ (dotted). (c) Time evolution of $\max_zY_-$ (solid red), $\max_z(H-Y_+)$ (solid blue) and $\max_z(|\tilde{\tau}_w|-B)$ (solid black) for the simulation in (a). Also shown are plots of $\max_zY_{-}$ (dot-dashed red) and $\max_z(|\tilde{\tau}_w|-B)$ (dot-dashed black) for the surfactant-free viscoplastic simulation $(B=0.04,M=0)$.}}
\end{figure*}

Figure \ref{fig:TFevolution}(a) shows snapshots from a sample numerical solution of the thin-film system. At $\tilde{t}=0$, the initial perturbation amplitude, $A$, is large enough that the layer is yielded in a region in the centre of the domain with a fully-yielded region at the base of the layer. At very early times, this yielded region spreads to cover the whole domain by $\tilde{t}=5$. In this very-early-time period, $\Gamma_z$ remains small, so the dynamics are essentially uninfluenced by the presence of surfactant. The behaviour is qualitatively the same as in the early-time yielding period described previously for the surfactant-free problem \citep{shemilt_2022_surface}. Figure \ref{fig:TFevolution}(b) shows that there is a delay in the initial growth of the instability compared to a Newtonian ($B=0$) simulation, and that the delay is approximately equal in the surfactant-free viscoplastic simulation. After this early-time period, however, Figure \ref{fig:TFevolution}(b) shows that the growth rate of the instability is reduced by the presence of surfactant. 

Between $\tilde{t}=5$ and $\tilde{t}=80$ (figure \ref{fig:TFevolution}a), significant gradients in surfactant concentration develop, and at $\tilde{t}\approx80$ a fully yielded region at the interface, where $M\Gamma_z$ exceeds $B$, appears near the centre of the domain. This yielded region grows and has extended across the whole interface by $\tilde{t}=500$. During the intermediate period, $80<\tilde{t}<500$, the two lateral edges of the upper yielded region propagate towards the side boundaries; these coincide with the location of two gradually steepening travelling wave fronts in $\Gamma_z$. At $\tilde{t}=130$, the right-travelling wave in $\Gamma_z$ near $z=4$ resembles a discontinuous shock wave. The sudden decrease in $\Gamma_z$ across the shock results in a rise in $Y_-$ just ahead of the shock where the Marangoni force is weaker. Similarly, at $\tilde{t}=250$, a shock-like discontinuity has developed in $\Gamma_z$ in the left-travelling wave as it approaches $z=0$, and a small rise in $Y_-$ can be observed ahead of the wave. 

Once we observe that a shock has developed in $\Gamma_z$, we can use \eqref{surftransportTF} to derive a Rankine-Hugoniot condition (see Appendix \ref{app:propagation}), which provides the relation 
\begin{equation}
    u_s = \frac{\left[(\tilde{w}_s\Gamma)_z\right]_-^+}{\left[\Gamma_z\right]_-^+}\label{mainRHcondition}
\end{equation}
for the shock propagation speed, $u_s$, in terms of the sizes of jumps in $\Gamma_z$ and $(\tilde{w}_s\Gamma)_z$ across the discontinuity. We have already noted how the jump in $\Gamma_z$ across the shock affects the yielding behaviour on either side, but \eqref{mainRHcondition} also shows that the yielding behaviour, via $\tilde{w}_s$, influences the shock propagation, highlighting the coupling between rheology and surfactant transport. Although our numerical method does not actively track the shock location, the speed of shock propagation observed in the simulations agrees well with $u_s$ calculated via \eqref{mainRHcondition} (data not shown here), providing evidence that the numerics accurately capture the behaviour around the shock.

Throughout the intermediate-time period described above, the regions ahead of the shock waves exhibit yield type II (surface pseudo-plug) while the central region behind the shocks exhibits yield type I (internal pseudo-plug). At $\tilde{t}=80$, $\tilde{t}=100$ and $\tilde{t}=130$, at both side boundaries the solution in $(H\tilde{p}_z$/B,$M\Gamma_z/B)$-space approaches $(H\tilde{p}_z/B,M\Gamma_z/B)=(0,-1)$. This indicates that near the side boundaries, capillary forces are small and Marangoni forces dominate. Unlike in the clean problem, where $Y_-\rightarrow0$ as $z\rightarrow0$ or $z\rightarrow L$, here due to the presence of Marangoni forces, $Y_-$ is not necessarily zero at the side boundaries. The value of $Y_-$ in the limits $z\rightarrow0$ or $z\rightarrow L$ during this period is given by $\lim_{z\rightarrow0,L}\left(H-M\Gamma_{zz}/\tilde{p}_{zz}\right)$, which can be seen to take a positive value in Fig \ref{fig:TFevolution}(a) for $80\leq \tilde{t}\leq130$. 

Once the shock waves have propagated to the side boundaries, the evolution enters a late-time regime where the upper and lower yielded regions extend across the whole layer while getting gradually smaller, indicating that the layer is rigidifying. Both $Y_-$ and $H-Y_+$ tend towards zero at a rate proportional to $1/t$ (figure \ref{fig:TFevolution}c). It can also be seen that in $(H\tilde{p}_z/B,M\Gamma_z/B)$-space (figure \ref{fig:TFevolution}a, $\tilde{t}=500,2000$), the whole solution is close to the line $M\Gamma_z+\frac{1}{2}H\tilde{p}_z=0$. This indicates that $H-Y_+\approx Y_-$, and so $\tilde{w}_s\approx0$ \eqref{wsTF}, meaning the interface is approximately immobilised at late times. As the layer approaches its final static shape, all points in the domain converge towards $(H\tilde{p}_z/B,M\Gamma_z/B)=(-2,1)$ (figure \ref{fig:TFevolution}a, $\tilde{t}=2000$). Hence, when the layer reaches its final static shape, the capillary stress is twice as strong as the Marangoni stress and they act in opposite directions, with the resultant magnitude of stress exactly equal to $B$. From this, we can deduce that at late times the layer approaches a marginally-yielded static shape, $H\rightarrow H_0(z;B)$ as $\tilde{t}\rightarrow\infty$, which satisfies 
\refstepcounter{equation}
$$
    H_0(H_{0,z}+H_{0,zzz}) = 2B,\label{H0eqn}
    \eqno{(\theequation{\mathit{a}})}
$$
whilst $M\Gamma\rightarrow M\Gamma_0$ as $\tilde{t}\rightarrow\infty$ where $\Gamma_0$ has a linear profile with slope $B/M$ and mass $L$ over $0<z<L$,
$$
    M\Gamma_0 = M - \frac{BL}{2} + Bz.\label{MaG0eqn}
    \eqno{(\theequation{\mathit{b}})}
$$
For the solution (\ref{H0eqn}b) to be valid, it must have $\Gamma_0\geq0$ everywhere, or specifically $2M\geq BL$. Indeed, we observe that in simulations with $2M<BL$, $H$ and $\Gamma$ do not approach $H_0$ and $\Gamma_0$ at late times, but rather approach different static solutions: we will discuss this in more detail in \S \ref{sec:TFstability}. Until then, we will focus attention on the case where $M$ is sufficiently large that a solution of \eqref{H0eqn} is approached at late times. 

In the surfactant-free problem \citep{shemilt_2022_surface}, the late-time static solution for $H$ also satisfies the ODE (\ref{H0eqn}a) but with $2B$ replaced by $B$. Hence, the presence of sufficiently strong surfactant effectively doubles the capillary Bingham number in relation to the layer's final shape. In keeping with this observation, figure \ref{fig:TFevolution}(b) indicates the decreased late-time height of the layer compared to the clean viscoplastic problem. Figure \ref{fig:TFevolution}(c) shows the evolution of $\max_z(|\tilde{\tau}_w|)$, the maximum value of the shear stress exerted on the tube wall, showing that it peaks around $\tilde{t}\approx100$ and then approaches $B$ at late times. The comparison with the surfactant-free simulation in the same figure shows that introducing surfactant reduces the peak in the wall stress.

\subsection{\label{sec:latetimeasymptotics}Late-time dynamics of a thin film}

In order to predict the late-time behaviour of the layer, we propose an asymptotic solution for $\tilde{t}\gg1$. Inspired by observations from numerical simulations such as figure \ref{fig:TFevolution}, we make expansions of the form

\begin{equation}
\left.\begin{array}{c}
\displaystyle
         H = H_0 + \frac{H_1}{Bt}+\dots,\quad \Gamma = \Gamma_0 + \frac{\Gamma_1}{Bt}+\dots,\quad
        Y_+ = H_0 + \frac{Y_{+,1}}{Bt}+\dots,\\[14pt]
        \color{black}
        \displaystyle
        Y_- = \frac{Y_{-,1}}{Bt}+\dots,
        \quad
        \p_zp = \p_zp_0 + \frac{\p_zp_1}{Bt}+\dots = -\frac{2B}{H_0} - \frac{(\p_z+\p_z^3)H_1}{Bt} + \dots, 
        \end{array}\right\}
        \label{latetimeexpansions}
\end{equation}
\color{black}
where $\Gamma_0$ is given by (\ref{H0eqn}b) and $H_0$ is a solution to (\ref{H0eqn}a). For all of the analysis in this section, we will assume that $2M\geq BL$, so (\ref{H0eqn}b) is a valid solution. We will also assume that $Y_{-,1}>0$ and $Y_{+,1}<0$ for $0<z<L$, since we observe in numerical simulations (e.g. figure \ref{fig:TFevolution}) that the whole layer exhibits yielding of type I when it is in this late-time regime. We will subsequently confirm that the resulting solution is consistent with simulations. Substituting \eqref{latetimeexpansions} into \eqref{Ydefns}-\eqref{qdefnTF} gives
\refstepcounter{equation}
$$
    H_1 = \partial_z\left(Y_{-,1}^2\right),\quad Y_{+,1} = H_1 - \frac{H_0}{2B}M\partial_z\Gamma_{1},\quad Y_{-,1} = Y_{+,1} + \frac{H_0^2}{2B}(\partial_z+\partial_z^3)H_1.
    \eqno{(\theequation{\mathit{a}-{c}})}
    \label{latetimeeqns}
$$
Inserting \eqref{latetimeexpansions} into the surface velocity \eqref{wsTF} gives, to leading order,
\begin{equation}
    \tilde{w}_s \sim \frac{1}{B^2t^2}W_1, \quad \mathrm{where}\quad W_1 = \frac{BY_{-,1}^2}{H_0}-\frac{H_0}{4B}(M\p_z\Gamma_{1})^2,
\end{equation}
and the transport equation \eqref{surftransportTF} gives
\begin{equation}
    B\Gamma_1 = \partial_z\left(\Gamma_0W_1\right).
    \label{latetimeG1eqn}
\end{equation}
The boundary conditions \eqref{TFsideBCs} imply
\begin{equation}
    \partial_zH_0 = \partial_zH_1 = Y_{-,1} = W_1 = 0 \quad\mathrm{at}\quad z= \{0,L\},
    \label{latesideBCs}
\end{equation}
and mass conservation implies
\begin{equation}
    \int_0^LH_0\,\mathrm{d}z = L,\quad \int_0^LH_1\,\mathrm{d}z = 0, \quad\int_0^L\Gamma_1\,\mathrm{d}z=0.
    \label{latemassBCs}
\end{equation}
We solve the equations (\ref{H0eqn}a) and \eqref{latetimeeqns}-\eqref{latetimeG1eqn} numerically subject to \eqref{latesideBCs} and \eqref{latemassBCs}.

\begin{figure*}
    \centering
    \includegraphics[width=\textwidth]{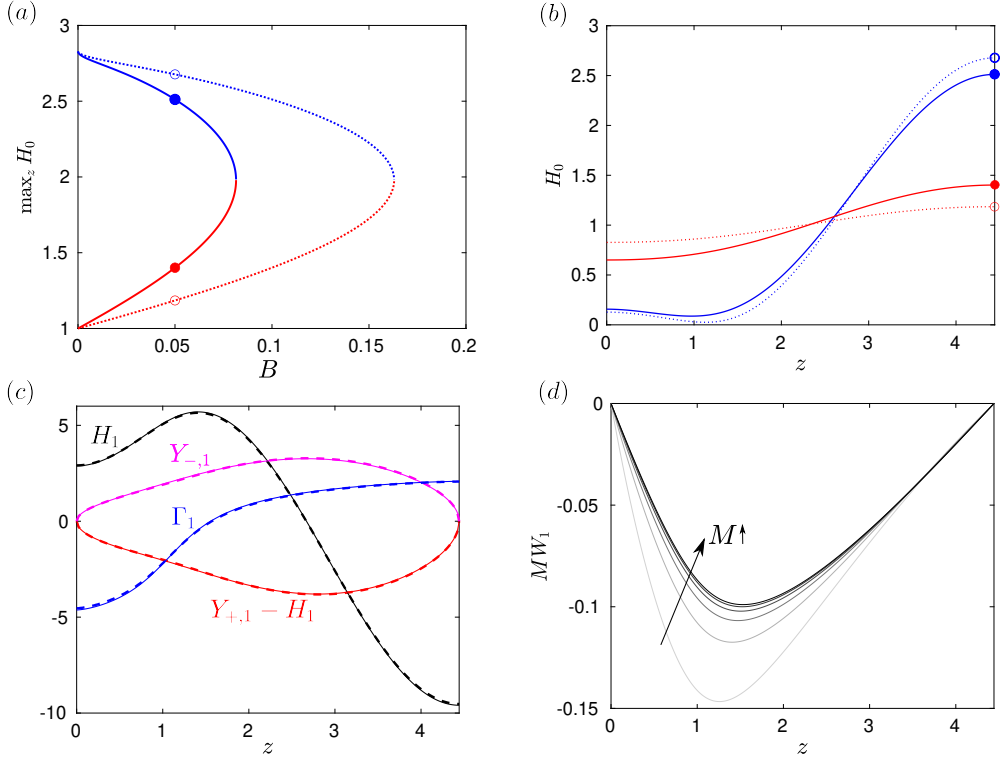}
    \caption{(a) Plot of $\max_zH_0$ for all the solutions to (\ref{H0eqn}a) (solid), which are static, marginally-yielded solutions to the thin-film equations with surfactant. For comparison, the equivalent plot for the static solutions for the surfactant-free problem, which were computed in \citet{shemilt_2022_surface}, is also shown (dotted). The four coloured markers in (a) correspond to the example solutions shown in (b). All solutions in (b) have $B=0.05$ but the dotted ones are the solutions for the surfactant-free problem. (c) $O(1/t)$ terms in the late-time expansion \eqref{latetimeexpansions} for a solution with $B=0.05$ and $M=0.5$, showing $H_1$ (solid black), $Y_{-,1}$ (solid magenta), $Y_{+,1}-H_1$ (solid red) and $\Gamma_1$ (solid blue). These are compared to corresponding quantities from a numerical solutions of the thin-film equations \eqref{Ydefns}-\eqref{TFsideBCs} at $\tilde{t}=10^4$, specifically $[H(z,\tilde{t}=10^4)-H_0(z)]B\tilde{t}$ (dashed black), $Y_-(z,\tilde{t}=10^4)B\tilde{t}$ (dashed magenta), $[Y_+(z,\tilde{t}=10^4)-H(z,\tilde{t}=10^4)]B\tilde{t}$ (dashed red) and $[\Gamma(z,\tilde{t}=10^4)-\Gamma_0(z)]B\tilde{t}$ (dashed cyan) where $\Gamma_0$ is defined in (\ref{H0eqn}b). 
    (d) Leading-order surface velocity, $W_1$, scaled by $M$, for $B=0.05$ and $M=\{0.25,0.5,1,2,4,8\}$.}
    \label{fig:latetime}
\end{figure*}

There are two branches of solutions for $H_0$, as shown in figure \ref{fig:latetime}(a), which are the same set of static solutions as computed in the surfactant-free problem \citep{shemilt_2022_surface} but with each corresponding $B$ halved. The upper-branch solutions are strongly deformed, marginally-yielded states that the layer approaches at late times. Figure \ref{fig:latetime}(b) illustrates how the deformation and peak height of these solutions is reduced when surfactant is present compared to when the interface is clean, and figure \ref{fig:latetime}(a) quantifies this effect for varying $B$. The lower-branch solutions in figure \ref{fig:latetime}(a) are unstable near-flat marginally-yielded states, which in the clean problem were shown to approximate the minimum initial perturbation required to trigger instability \citep{shemilt_2022_surface}. We will show in \S \ref{sec:TFstability} that the lower-branch solutions for the surfactant-laden problem have the same significance when $M$ is sufficiently large. Figures \ref{fig:latetime}(a) and \ref{fig:latetime}(b) quantify the increased deformation in the lower-branch solutions when surfactant is present compared to when it is not.

Figure \ref{fig:latetime}(c) shows the $O(1/t)$ terms in an example late-time asymptotic solution. There is close agreement between these asymptotics and the numerical solution of the thin-film equations at $\tilde{t}=10^4$. Also, figure \ref{fig:latetime}(c) suggests that $Y_{-,1}\approx Y_{+,1}-H_1$, so at late times the near-wall and near-interface yielded regions have approximately the same size. This implies that the surface velocity, $\tilde{w}_s$, is approximately zero. 
We investigate this further by plotting the leading-order surface velocity, $W_1$, for various $M$ in figure \ref{fig:latetime}(d). As $M$ is increased, $MW_1$ approaches a fixed curve, indicating that $W_1$ is approaching zero at a rate proportional to $O(1/M)$. It also shows that $W_1<0$ in all cases, indicating that at late times surfactant typically induces a reverse flow at the interface. However, at sufficiently large $M$, the surface velocity is very small and so the interface is essentially immobilised as the layer approaches its late-time configuration. \textcolor{black}{By immobilisation of the interface, we mean that there is no axial surface velocity, but the free surface can still deform and the layer thickness can still evolve.}

\subsection{Effect of varying the Marangoni number on stability and dynamics}\label{sec:TFstability}

\begin{figure*}
    \centering
    \includegraphics[width=0.9\textwidth]{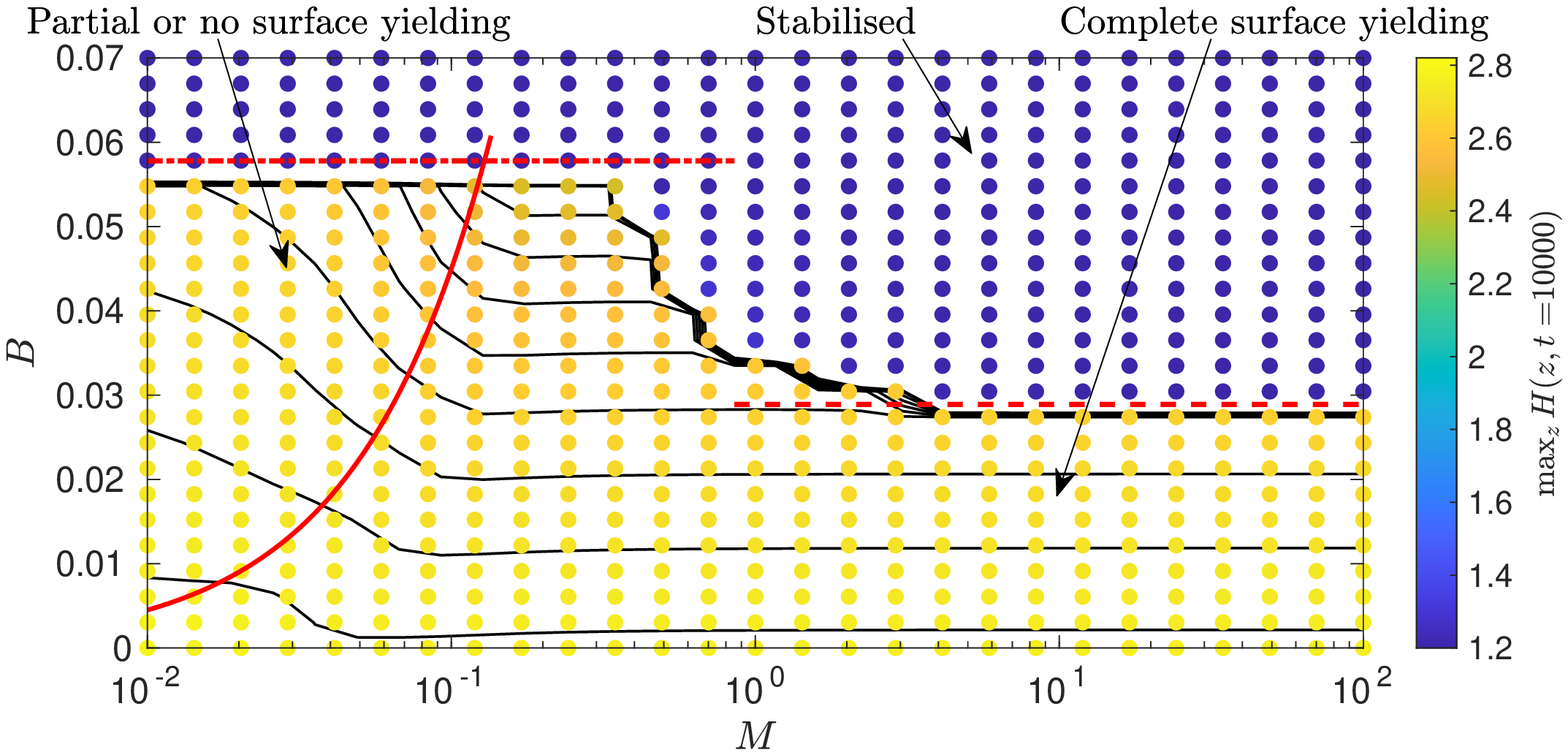}
    \caption{Data from thin-film simulations with $A=0.2$ and various $B$ and $M$. Each coloured point corresponds to one simulation with the colour indicating the final peak height. Contours interpolated from the same data are also plotted in black, which are evenly spaced and in the range $2.4\leq\max_zH\leq2.8$. The red lines are $2M=BL$ (solid), $B=B_m(A=0.2)\approx0.0289$ (dashed) and $B=2B_m(A=0.2)\approx0.0578$ (dot-dashed). }
    \label{fig:BvsMaTF}
\end{figure*}

To systematically assess the effect of surfactant on the stability and dynamics of the layer, we have run a large set of numerical simulations for various values of $B$ and $M$, with a fixed initial perturbation to the layer height, $A=0.2$. Figure \ref{fig:BvsMaTF} shows the resulting data for the final peak height, $\max_zH(z,\tilde{t}=10^4)$, which have several characteristics of note. 

There is a sharp stability boundary in the data in figure \ref{fig:BvsMaTF}. There is a critical value of $B$, which we call $B_{c}$, such that for $B<B_{c}$ there is instability and significant deformation of the layer, whilst for $B>B_{c}$ there is minimal or no deformation. This stabilisation at high $B$ was identified for the surfactant-free problem \citep{shemilt_2022_surface}, but it can be seen from figure \ref{fig:BvsMaTF} that $B_{c}$ also depends on $M$. For small $M$, $B_{c}$ recovers its value for the surfactant-free case but, as $M$ is increased, $B_{c}$ decreases towards a different constant value at high $M$. It is evident that $B_{c}$ also depends on the amplitude of the initial perturbation given to the layer, $A$\textcolor{black}{, as we discuss in detail in \S\ref{sec:Adependence} below.} In \citet{shemilt_2022_surface}, we showed that $B_{c}$ can be predicted accurately for a large range of $A$ using the marginally-yielded static solutions, $H_0(z;B)$. Here, we find that the same approach can be used to predict $B_{c}$ for large $M$ as well as small $M$. If we define $B_m(A)$ as the value of $B$ such that the solution $H_0$ to equation (\ref{H0eqn}a) satisfies $1-H_0(z=0;B)=A$, then we find $B_m(0.2)\approx0.0289$ which can be seen in figure \ref{fig:BvsMaTF} to accurately predict $B_{c}$ at large $M$. To approximate the stability boundary at small $M$, we can use the prediction for the surfactant-free problem, which is simply $B_{c}\approx 2B_m\approx0.0578$, since the function $H_0$ is the same but the corresponding capillary Bingham number for that $H_0$ is doubled. Figure \ref{fig:BvsMaTF} shows that this predicts the stability boundary well at small $M$. 

To understand why the prediction for large $M$ works, first note that, in the surfactant-free problem \citep{shemilt_2022_surface}, the lower-branch static solutions from figure \ref{fig:latetime}(a) correspond to the minimum amplitude of perturbation required to make the layer initially yield, as long as we now also have that $M\Gamma_z\approx B$. For large $M$, surfactant is strong enough that a thin fully-yielded region adjacent to the interface develops rapidly before any significant deformation of the layer occurs, so after a rapid adjustment from the initial conditions at early times, we do have $M\Gamma_z\approx B$ subsequently. Then, whether instability occurs depends only on whether the initial perturbation to the layer height, parameterised by $A$, was larger than the minimum required to yield, which is represented by the lower-branch static solution for a given $B$. Or, equivalently, if $A$ is fixed as in figure \ref{fig:BvsMaTF}, then instability occurs only if $B<B_m(A)$.

In the limit of very strong surfactant, we can derive a simplified evolution equation by exploiting $M\gg1$ as a large parameter (see Appendix \ref{app:TFlargeM}). In this large-$M$ asymptotic theory, we find that the surface velocity is zero to leading order, so the interface is immobilised throughout the evolution. Additionally, we find that the motion is governed by the evolution equation,
\begin{equation}
    H_{\tilde{t}} + \left[\frac{1}{6}\tilde{p}_z\tilde{Y}^2\left(2\tilde{Y}-3H\right)\right]_z=0, \quad\mbox{where}\quad \tilde{Y} = \max\left(0,\frac{1}{2}H-\frac{B}{|\tilde{p}_z|}\right).
    \label{largeMevoleqn}
\end{equation}
If we replace $B$ with $\frac{1}{2}B$ and $\tilde{t}$ with $\frac{1}{4}\tilde{t}$ in \eqref{largeMevoleqn}, then we exactly recover the evolution equation for the surfactant-free problem (equation (2.27) in \citet{shemilt_2022_surface}). Therefore, solutions for the dynamics with very strong surfactant are the same as solutions for the dynamics without surfactant, but with the capillary Bingham number doubled and time slowed by a factor of four. The four-fold increase in the time scale has been established previously for Newtonian fluids \citep{otis_role_1993}, but the effective doubling of the yield stress by strong surfactant is (we believe) a new phenomenon. 
We have already seen this doubling of $B$ in relation to the final shape of the layer in (\ref{H0eqn}a) and in the value for $B_c$ in figure \ref{fig:BvsMaTF}, as described above. However, this theory indicates that for $M\gg1$ the doubling of $B$ holds for the entire dynamics, except for a very short period with time scale $O(1/M)$ at the beginning of the evolution when the initially uniform surfactant profile rapidly adjusts to a configuration consistent with the asymptotic theory.

Figure \ref{fig:BvsMaTF} also indicates that, within the unstable region of parameter space, i.e. $B<B_{c}$, there are two qualitatively different behaviours. At sufficiently large $M$, there is a region where the final peak height is effectively independent of $M$. These are the simulations which obey the late-time behaviour described in \S \ref{sec:latetimeasymptotics}, and so their final shape is the marginally-yielded static shape $H_0$, a solution to (\ref{H0eqn}a), which has no $M$-dependence. 
In contrast, for sufficiently small $M$, not all of the fluid adjacent to the interface fully yields at late times, and so the layer does not enter the late-time regime described in \S \ref{sec:latetimeasymptotics} but instead approaches a different final static shape. It can be seen in figure \ref{fig:BvsMaTF} that for small $M$, the final peak height and, hence, the final shape, depend on both $M$ and $B$. \textcolor{black}{These results are consistent with the observation that the late-time solution \eqref{H0eqn} exists only for $2M\geq BL$: figure \ref{fig:BvsMaTF} shows there is $M$-dependence in the final layer shape when $2M<BL$. There is, however, also a small band of parameter values in figure \ref{fig:BvsMaTF} where there is still $M$-dependence in the final shape despite $2M\geq BL$. Although our results suggest that the final shape of the layer in these cases does depend on $M$, we cannot rule out the possibility that if simulations were run to much longer times some of these simulations may eventually approach the $M$-independent final shape, $H_0$.}

\begin{figure*}
    \centering
    \includegraphics[width=\textwidth]{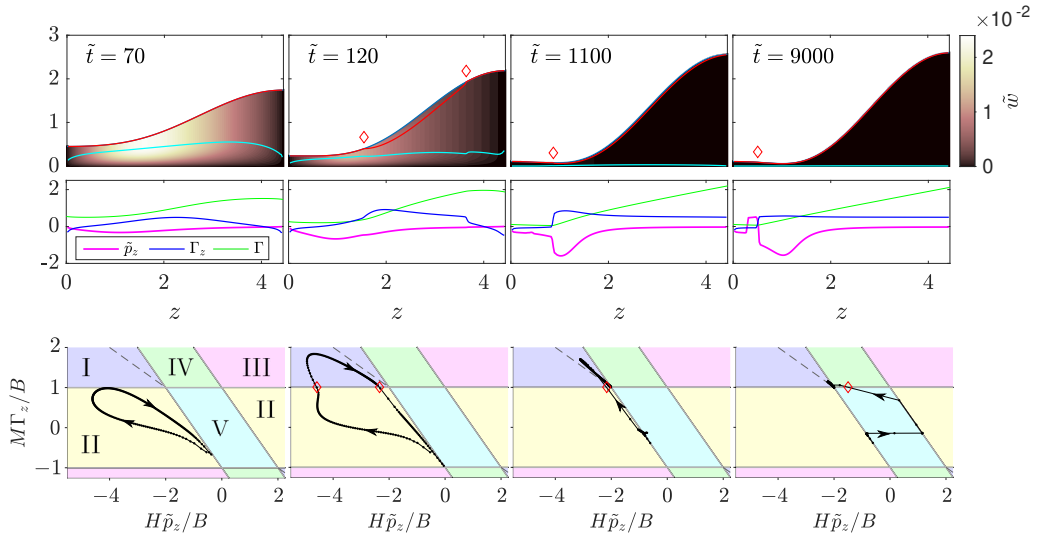}
    \caption{Snapshots from a thin-film simulation with $B=0.04$, $M=0.08$, $A=0.2$, at $\tilde{t}=\{70,120,1100,9000\}$. The upper row of panels shows the layer height evolving, with $Y_-$ (cyan) and $Y_+$ (red), and the thin-film axial velocity, $\tilde{w}$, represented by the colour map; the middle row shows $\Gamma$ (green), $\Gamma_z$ (blue) and $\tilde{p}_z$ (magenta); and the lower row shows shows the solution in $(H\tilde{p}_z/B,M\Gamma_z/B)$-space, with the black dots corresponding to evenly spaced points along $0<z<L$ and the arrows indicating the direction of increasing $z$. \textcolor{black}{Red diamonds on the first and third panels mark the boundaries of the region where there is yielding at the interface.}}
    \label{fig:smallMasim}
\end{figure*}

Figure \ref{fig:smallMasim} shows an example numerical simulation in this small $M$ region. In this simulation, there is a region near $z=0$ that exhibits yielding of type II throughout the evolution, even at late times, so there is only partial yielding adjacent to the interface. For this simulation, $B=0.04$ and $M=0.08$, so $2M<B L$. This means that the layer cannot enter the late-time regime described in \S \ref{sec:latetimeasymptotics}, as then (\ref{H0eqn}b) would require $\Gamma$ to be negative. In figure \ref{fig:BvsMaTF}, the boundary between the region with complete surface yielding (simulations that enter the late-time regime from \S \ref{sec:latetimeasymptotics}) and the small $M$ region with no or partial surface yielding occurs close to the line $2M=B L$. Physically, we can interpret the behaviour at small $M$ as the surfactant being too weak for Marangoni effects to be able to fully yield the whole interface. In figure \ref{fig:smallMasim} a large portion of the domain does exhibit interface-adjacent yielding, but if $M$ was decreased further this portion would become smaller and eventually as $M\rightarrow0$ no fully-yielded region would exist near the interface, consistent with the behaviour in the surfactant-free case. We note that at late times in the simulation in figure \ref{fig:smallMasim}, there is complex behaviour near $z=0$ with a small region developing where $\tilde{p}_z$ changes sign. We find this type of behaviour to be typical in simulations with very small $M$, but since it occurs at late times when the layer is already near-static, it generally has minimal impact on the global dynamics. Finally, figure \ref{fig:BvsMaTF} demonstrates that surfactant has an appreciable impact on the final configuration of the thin film for $M=O(1)$, corresponding (from \eqref{BMadefns}) to a negligible ($O(\epsilon^2)$) reduction in surface tension relative to its mean value. In this thin-film limit, surfactant operates solely through powerful Marangoni effects. 

\color{black}

\subsection{Dependence on the amplitude of initial perturbation}\label{sec:Adependence}

\begin{figure}
    \centering
    \includegraphics[width = \textwidth]{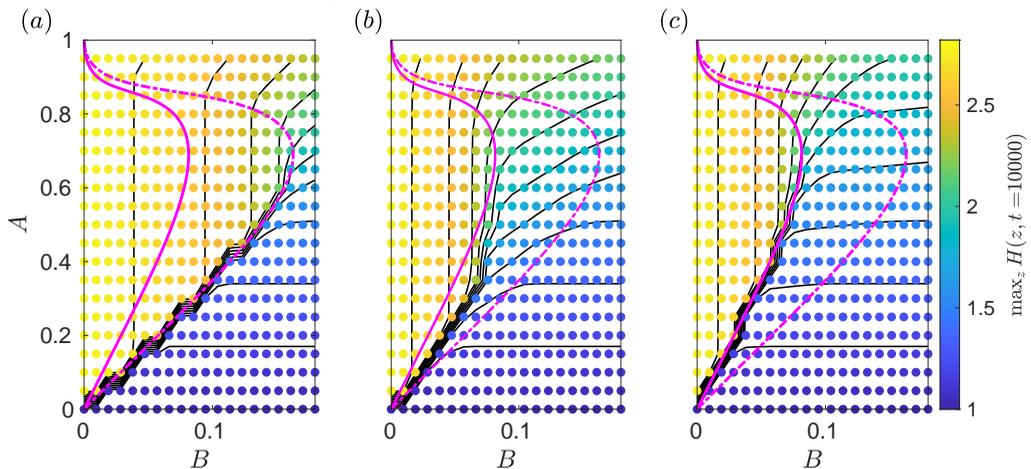}
    \caption{\textcolor{black}{Data from thin-film numerical simulations with (a) $M=0$, (b) $M=0.6$ and (c) $M=6$. Each coloured dot corresponds to one simulation, with the given values of $A$ and $B$, where the colour indicates the final maximum height of the layer. The same data for $\max_zH(z,t=10^4)$ is linearly interpolated and plotted as black contour lines. The two magenta curves on each panel are $B=B_m(A)$ (solid) and $B=2B_m(A)$ (dot-dashed). }}
    \label{fig:AvsB}
\end{figure}

When $B>0$, the instability is nonlinear and the dynamics of the layer, including its final shape, have some dependence on the size of initial perturbation to the free surface, $A$. For the surfactant-free problem, \citet{shemilt_2022_surface} showed that the late-time static solutions (solutions to equation (\ref{H0eqn}a) but with $2B$ replaced by $B$) can be used both to predict the minimum $A$ required to trigger instability, referred to as $A_c(B)$, and to identify a large region of parameter space where the final shape is independent of $A$.  Figure \ref{fig:AvsB}(a) illustrates how, in the surfactant-free case, the curve $B = 2B_m(A)$ approximates closely the boundary of the region in $(B,A)$-space where $\max_zH(z,t=10^4)$ is independent of $A$. This boundary includes the sharp stability boundary separating simulations where minimal or no growth occurred from the simulations where there is significant growth. As above, $B_m(A)$ is defined as the value of $B$ such that $1-H_0(z=0;B) = A$, where $H_0$ are solutions to (\ref{H0eqn}a). 

Figures \ref{fig:AvsB}(b) and \ref{fig:AvsB}(c) show how introducing surfactant alters this $A$-dependence, particularly how $A_c$ is increased for larger values of $M$. For $M=0.6$ (figure \ref{fig:AvsB}b), $A_c$ is increased for each value of $B$ compared to the surfactant-free case. When surfactant strength is increased again to $M=6$ (figure \ref{fig:AvsB}c), $A_c$ is further increased and, analogously to the surfactant-free case, the curve $B=B_m(A)$ predicts the entire boundary of the region where the final shape is independent of $A$, including predicting $A_c$. The lower branch of static solutions in figure \ref{fig:latetime}(a) correspond to the values of $B_m$ that predict $A_c$ in figure \ref{fig:AvsB}(c), illustrating that these lower-branch solutions correspond to the minimal initial perturbation to the free surface required to trigger instability when surfactant is strong. The accuracy of $B=B_m(A)$ in predicting the whole boundary of the $A$-independent region in figure \ref{fig:AvsB}(c) provides further evidence of the effective doubling of $B$ by introducing strong surfactant compared to the surfactant-free case, and highlights that the static solutions, $H_0(z;B)$, can provide significant insight into the stability and dynamics of a layer with strong surfactant. Figure \ref{fig:AvsB} also illustrates that, while there is non-trivial dependence of $A_c$ on $M$, accurate upper and lower bounds for $A_c$ for any $M$ are provided by the curves $B=B_m(A)$ and $B=2B_m(A)$, respectively.

\color{black}

\section{\label{sec:LW}Thick films and liquid plug formation}

Whilst the relative simplicity of the thin-film equations allows for more detailed analysis, liquid plug formation cannot occur in that theory. In order to assess the effect of surfactant on the dynamics leading to plug formation, we now consider the long-wave equations \eqref{LWevoleqn}-\eqref{LWsideBCs}, which model the evolution of layers which are not thin. The layer thickness is described by the parameter $\epsilon$, which is the ratio of the average thickness to the tube radius after a small adjustment to account for the change in volume induced by having a finite amplitude initial perturbation (see \eqref{LWICs}).

The minimum volume of fluid required to form a liquid plug corresponds to a layer thickness of $\epsilon\approx0.107$ for the length of domain we are using \citep{everett_model_1972}. \citet{gauglitz_extended_1988} conducted the first numerical simulations of the Newtonian problem without surfactant, which predicted a slightly larger critical thickness, $\epsilon_{\text{crit}}\approx0.12$, due in part to the restriction of only being able to run simulations to a relatively short, finite time. 
Viscoplastic rheology can significantly increase $\epsilon_{\text{crit}}$ when the capillary Bingham number is sufficiently high \citep{shemilt_2022_surface}. Here, we investigate how $\epsilon_{\text{crit}}$ is affected by the additional presence of surfactant. 

We run all simulations to time $\tilde{t}=10^4$, which is an order of magnitude longer than in \citet{shemilt_2022_surface}, in order to capture the dynamics around plug formation as accurately as possible. Following the approach of many previous studies, we stop the simulations early if $\min_zR \leq 0.3$, or equivalently if $\max_zH\geq 0.7/\epsilon$, as this has been found to indicate that a plug is imminently about to form \citep{halpern_fluid-elastic_1992,halpern_surfactant_1993,halpern_effect_2010,shemilt_2022_surface}. We then identify the time at which the simulation is stopped as the plugging time, $t_p$. 
We will describe the thick-film results in this section using the unscaled capillary Bingham and Marangoni numbers, $\B$ and $\Ma$, defined in the long-wave theory \eqref{BMadefns}, rather than the scaled versions that arise in the thin-film theory. One can convert back to the scaled, thin-film parameters by dividing $\B$ and $\Ma$ by $\epsilon^{2}$.

\subsection{Example evolution showing plug formation\label{sec:LWexamples}} 

\begin{figure*}

\centering
\includegraphics[width=\textwidth]{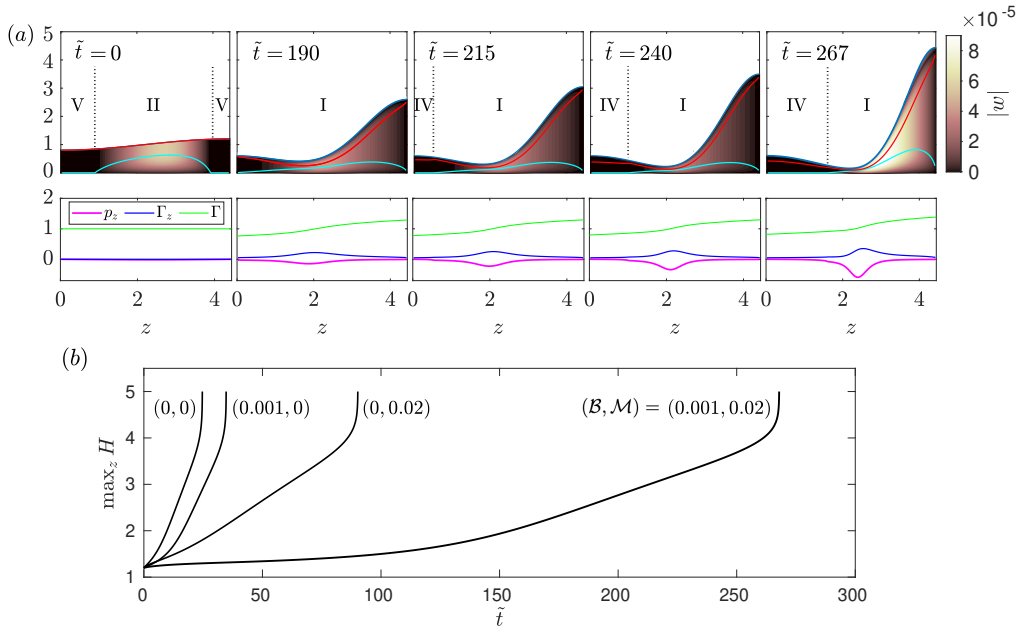}
\caption{(a) Numerical solution of the long-wave equations \eqref{Psipmdefn}-\eqref{LWsideBCs} with $A=0.2$, $\B=0.001$, $\mathcal{M}=0.02$ and $\epsilon=0.14$, showing the transition towards plug formation. The upper panel shows the evolution of the layer height, with $Y_-$ (cyan) and $Y_+$ (red) also shown, and the colour corresponding to the magnitude of the axial velocity, $|w|$. The lower panel shows $p_z$ (magenta), $\Gamma_z$ (blue) and $\Gamma$ (green). (b) Time evolution of $\max_zH$ for the same simulation, compared to $\max_zH$ from simulations with $(\B,\Ma)=\{(0,0),(0.001,0),(0,0.02),(0.001,0.02)\}$, showing the combined delay to plug formation by surfactant and yield stress.}
\label{fig:LWsimulation}
\end{figure*}

Figure \ref{fig:LWsimulation}(a) shows a numerical solution of the long-wave equations. The layer thickness is $\epsilon=0.14$, meaning that there is sufficient volume of fluid that a plug could form, and it can be seen in figure \ref{fig:LWsimulation}(b) that one does form at around $\tilde{t}\approx268$. Around $267\lesssim\Tilde{t}\lesssim 268$, the layer grows rapidly towards the centre of the tube, indicating that it is about to form a plug when the simulation is stopped. Introducing surfactant to a Newtonian layer with comparable thickness has been shown to increase the plugging time approximately four-fold \citep{ogrosky_linear_2021}. We find that adding surfactant to a viscoplastic layer can increase the plugging time by a much greater amount. In the example shown in figure \ref{fig:LWsimulation}(b), plug formation in the surfactant-laden viscoplastic film takes approximately eight times longer than when there is no surfactant, which in itself takes longer than plugging in the surfactant-free Newtonian simulation. 


Figure \ref{fig:LWsimulation}(a) shows that by $\tilde{t}=190$, the layer is exhibiting yielding of type I everywhere, similar to the typical thin-film behaviour seen in figure \ref{fig:TFevolution}(a). However, at later times, a region with type IV yielding, where the fluid is yielded near the interface but rigid adjacent to the wall, appears near $z=0$ and then expands towards the centre of the domain. The flow in this region is in the negative $z$-direction, opposing the flow in the rest of the layer. However, the velocity in this region is small so we expect any effect on the global dynamics to be minimal. This behaviour is similar to what is observed in the long-wave problem without surfactant, but in that case the region near $z=0$ is fully rigid \citep{shemilt_2022_surface}. Whilst we expect this long-wave theory to be a good approximation to the overall dynamics of the thick film, small details such as this behaviour near $z=0$ would need to be confirmed with full two-dimensional simulations. The more profound effect of surfactant that can be seen in figure \ref{fig:LWsimulation}(a) is the development of a sizeable fully yielded region near the interface all along the layer, where the shear is opposing the general flow and so delaying the formation of a plug.

\subsection{Thick-film dynamics at large Marangoni numbers}\label{sec:LWlargeM}

To explore the effect of increasing the surfactant strength on the evolution of a thick film, we propose an expansion for $\Ma\gg1$,
\begin{equation}
    R = \mathcal{R}_0 + \frac{1}{\Ma}\mathcal{R}_1+\dots,\quad \Gamma = \mathcal{G}_0 + \frac{1}{\Ma}\mathcal{G}_1+\dots,\quad w_s = \mathcal{W}_0 + \frac{1}{\Ma}\mathcal{W}_1+\dots.
    \label{LWlargeMexpansion}
\end{equation}
Since the surface stress must be finite, $|\p_z\sigma|=|\Ma\partial_z\Gamma|<\infty$, we must have $\partial_z\mathcal{G}_0=0$. Conservation of the total amount of surfactant in the long-wave theory then implies
\begin{equation}
    \mathcal{G}_0(t) = \frac{\int_0^L\mathcal{R}_0|_{t=0}\,\mathrm{d}z}{\int_0^L\mathcal{R}_0\,\mathrm{d}z}.
    \label{G0LW}
\end{equation}
The surfactant transport equation \eqref{surftransportLW} at leading order in $\Ma$, after rearranging and using \eqref{G0LW}, gives
\begin{equation}
        \mathcal{W}_0(z,t) = \frac{1}{\mathcal{R}_0}\int_0^z\mathcal{I}(\zeta,t)\,\mathrm{d}\zeta,\quad\mbox{where}\quad \mathcal{I} = \frac{\mathcal{R}_0\int_0^L\partial_t\mathcal{R}_{0}\,\mathrm{d}z-\p_t\mathcal{R}_{0}\int_0^L\mathcal{R}_0\,\mathrm{d}z}{\int_0^L\mathcal{R}_0\,\mathrm{d}z}.
        \label{w0LW}
\end{equation}
Therefore, unlike for the thin-film case (Appendix \ref{app:TFlargeM}), the surface velocity is not zero for large $\Ma$, in general. From \eqref{w0LW}, the surface velocity can be deduced from the difference between the local time derivative of the film thickness, $\p_t\mathcal{R}_0$, and the globally averaged rate of change of the thickness, via the integral of $\p_t\mathcal{R}_0$. Non-zero $\mathcal{W}_0$ leads to small gradients in surfactant concentration. The results \eqref{G0LW} and \eqref{w0LW} can be understood physically as follows: the surfactant is so strong that any local change in the shape of the interface induces a surface velocity that redistributes surfactant so that the concentration remains (almost) globally uniform, and if the total surface area of the interface has changed this will also induce a change in the mean concentration. We have appealed neither to the equation of state for surface tension \eqref{eqnofstate}, except that $\sigma'(\Gamma)\neq0$, nor to the liquid's rheology, to reach \eqref{G0LW} and \eqref{w0LW}. Hence, within long-wave theory, the results \eqref{G0LW} and \eqref{w0LW} are generic for any type of fluid and any equation of state that has $\sigma'(\Gamma)\neq0$. 

\begin{figure}
    \centering
    \includegraphics[width = \textwidth]{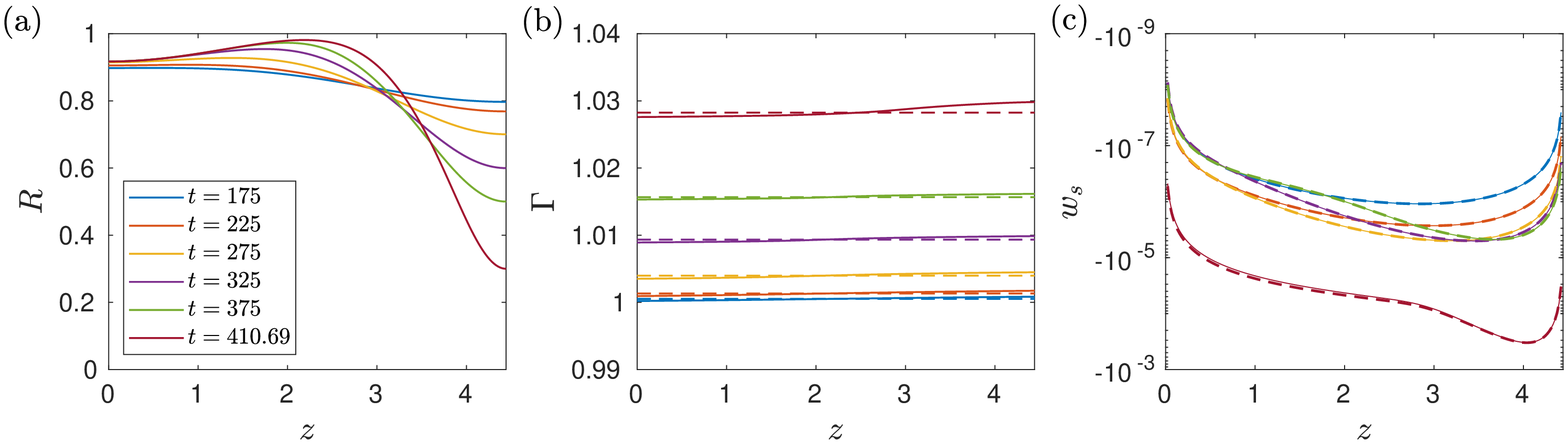}
    \caption{Results from a numerical solution of the long-wave equations with $\epsilon=0.14$, $\Ma=10$, $\B=0.001$ and $A=0.25$. (a) The interface position, $R$, shown at various time points. The last time point is $\tilde{t}=t_p=410.69$, when the simulation is stopped. (b) Surfactant concentration, $\Gamma$, (solid) at the same time points as in (a). These are compared to $\mathcal{G}_0(t)$ (dashed), the approximation from the large-$\Ma$ asymptotic theory, which is evaluated via \eqref{G0LW} using the numerical solutions from (a) as proxies for $\mathcal{R}_0$. (c) Surface velocity, $w_s$, (solid) at the same time points. These are also compared to the corresponding approximation from the large-$\Ma$ theory \eqref{w0LW}. }
    \label{fig:LWlargeMsim}
\end{figure}

Figure \ref{fig:LWlargeMsim} compares a solution of the full long-wave system \eqref{LWevoleqn}-\eqref{LWsideBCs} with $\Ma=10$ to the results \eqref{G0LW} and \eqref{w0LW}. It shows good agreement between $\Gamma$ and $w_s$ from the numerical solutions and the leading-order approximations \eqref{G0LW} and \eqref{w0LW}, despite the value of $\mathcal{M}$ not being extremely large. The spatial gradients in $\Gamma$ (figure \ref{fig:LWlargeMsim}b) are small in the numerical solution, and we expect these would become smaller if the surfactant strength was further increased, in line with the large-$\Ma$ theory. Figure \ref{fig:LWlargeMsim}(c) shows that the surface velocity induced as plug formation occurs is in the negative $z$-direction, and is strongest immediately before the simulation is stopped. \textcolor{black}{This is because there is a rapid decrease in the local surface area of the interface near $z=L$ during the fast growth before plug formation, which induces a surface velocity to redistribute surfactant across the layer.}

The results from figure \ref{fig:LWlargeMsim} suggest that the large-$\Ma$ theory can be used to better understand the behaviour of thick films when surfactant is strong. However, we have assumed a simple linear equation of state for surface tension \eqref{eqnofstate}, which limits how large we are able to make $\Ma$ in simulations while retaining accuracy. Figure \ref{fig:LWlargeMsim}(b) suggests that when the simulation is stopped, the increase in $\Gamma$ from its original value was around $3\%$. We have found that this is approximately independent of $\Ma$, so for values of $\Ma$ close to or larger than $\Ma\approx30$ we expect the theory to break down since $\sigma$ will approach zero as a plug forms. For this reason, $\Ma=10$ is the largest Marangoni that we have used in \textcolor{black}{the long-wave} simulations. A more complex nonlinear equation of state would have to be used to access higher values accurately.

\subsection{Critical layer thickness for plug formation\label{sec:LWcritical}}

\begin{figure*}
    \centering
    \includegraphics[width=\textwidth]{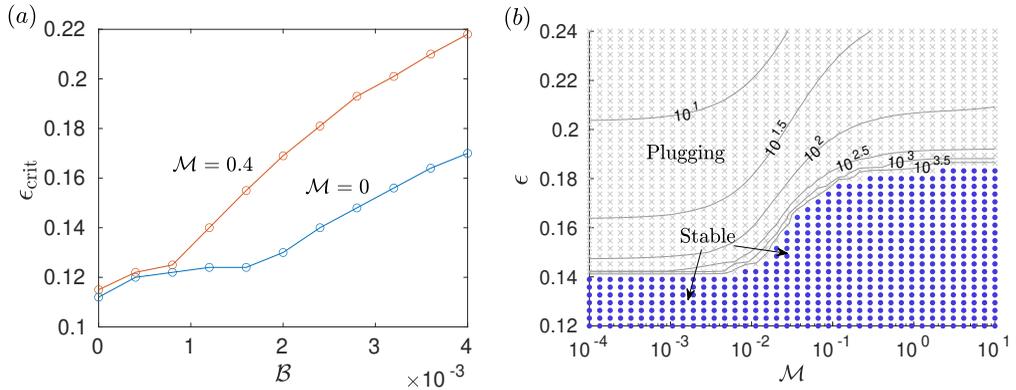}
    \caption
    {(a) Critical layer thickness, $\epsilon_{\text{crit}}$, as a function of $\mathcal{B}$, for a surfactant-free layer ($\mathcal{M}=0$) and a layer with surfactant ($\mathcal{M}=0.4$). All simulations have $A=0.25$. The value of $\epsilon_{\text{crit}}$ computed is such that a simulation with $\epsilon=\epsilon_{\text{crit}}+0.001$ forms a plug before $\tilde{t}=10^4$ and a simulation with $\epsilon=\epsilon_{\text{crit}}-0.001$ has not formed a plug by $\tilde{t}=10^4$. (b) Data from long-wave simulations at various values of $\epsilon$ and $\Ma$, with $A=0.25$ and $\B=0.0024$. Grey crosses indicate simulations where a liquid plug formed, while black dots indicate simulations where a plug did not form. Within the plugging region, the grey contours indicate the plugging time, $t_p$.}
    \label{fig:LWsweep}
\end{figure*}

Fig \ref{fig:LWsweep}(a) compares the computed critical layer thickness for plug formation to occur, $\epsilon_{\text{crit}}$, when surfactant is present to its value when the interface is clean. Both with and without surfactant, increasing $\B$ induces an increase in $\epsilon_{\text{crit}}$. The increase is small when $\B$ is small, but is significant at larger $\B$. When surfactant is present, the increase in $\epsilon_{\text{crit}}$ is initiated at $\B\approx10^{-3}$, compared to at $\B\approx2\times10^{-3}$ when the interface is clean. Beyond these values of $\B$, the increase is amplified by the presence of surfactant, such that the value of $\epsilon_{\text{crit}}$ is approximately $30\%$ larger than when the interface is clean. 

Figure \ref{fig:LWsweep}(b) illustrates the effect on the dynamics of varying $\Ma$ and $\epsilon$. It shows data from many simulations, some of which are stable and some form plugs. The boundary between the stable and unstable regions in the data can be identified as $\epsilon_{\text{crit}}$, so the dependence of the critical layer thickness on $\Ma$ can be seen. At low $\Ma$, the critical thickness for the surfactant-free problem is recovered, and as $\Ma$ is increased $\epsilon_{\text{crit}}$ increases to a significantly higher value for large $\Ma$. Therefore, for this value of $\B$, there is a range of layer thicknesses $0.14\lesssim\epsilon\lesssim0.185$ where increasing the surfactant strength sufficiently, without changing any other parameters, can suppress plug formation. This stabilisation of the system by increasing the surfactant strength resembles what was seen in the thin-film system (figure \ref{fig:BvsMaTF}). 
Again, surfactant has a powerful effect at low concentrations: the range $10^{-2}<\Ma<10^{-1}$, over which $\epsilon_{\mathrm{crit}}$ changes appreciably in figure \ref{fig:LWsweep}(b), corresponds to a maximum relative reduction of surface tension of between $1\%$ and $10\%$. Figure \ref{fig:LWsweep}(b) also quantifies how the plugging time increases as $\Ma$ is increased. We know increasing $\B$ extends the plugging time \citep{shemilt_2022_surface}, and figure \ref{fig:LWsweep}(b) shows it can be extended further again by introducing surfactant. However, the profound stabilisation effect by surfactant is not due mainly to slowing of the dynamics as it is in the Newtonian problem \citep{halpern_surfactant_1993}. Instead, figure \ref{fig:LWsweep}(b) highlights how the increase in $\epsilon_{\text{crit}}$ due to rigidification and stabilisation of the layer by yield stress is amplified by the presence of sufficiently strong surfactant.

The value of $\epsilon_{\text{crit}}$ and the dynamics leading to plug formation are necessarily dependent on the initial perturbation applied to the layer, which here is parameterised by $A$. With a larger initial perturbation, there can be more yielding at larger values of $\B$. However, we have found that varying the initial conditions does not qualitatively affect the results presented. The finite time to which we run simulations, $\tilde{t}=10^4$, will have some effect on the computed value of $\epsilon_{\text{crit}}$, but figure \ref{fig:LWsweep}(b) shows that the plugging time increases rapidly close to the computed value of $\epsilon_{\text{crit}}$, suggesting that there is good convergence to the true value of $\epsilon_{\text{crit}}$. 

\section{Discussion\label{sec:discussion}}

We have developed two models, using long-wave and thin-film theories, of the capillary instability of a viscoplastic layer coating a cylindrical tube where insoluble surfactant is present at the air-liquid interface. 
This flow can be strongly modified both by the strength of the yield stress, described by the capillary Bingham number, and the strength of the surfactant, described by the Marangoni number. We showed that the layer can exhibit five qualitatively different types of yielding, which are different combinations of fully-yielded shear flow, pseudo-plugs and rigid plugs, depending on the relative strengths of the capillary and Marangoni forces. Numerical simulations highlighted the complex dynamical transitions between these yielding types that can occur as the layer evolves. 

For thin layers, we quantified how introducing surfactant enhances the stabilising effect of the yield stress and induces an effective doubling of $B$ when $M$ is sufficiently large. Asymptotic analysis of the late-time behaviour in \S \ref{sec:latetimeasymptotics}, valid for moderate and large $M$, showed that the final static, marginally-yielded configuration of a surfactant-laden layer coincides with that of a surfactant-free layer but with $B$ exactly doubled. Static, marginally-yielded solutions were also shown to accurately predict the critical capillary Bingham number above which instability is suppressed, for both small and large Marangoni numbers (figure \ref{fig:BvsMaTF}). When $M$ is asymptotically large, the entire thin-film dynamics coincides with that of a surfactant-free layer but with time slowed by a factor of four and $B$ doubled. 

For thicker layers, we quantified how the known effect of yield stress in delaying or suppressing plug formation \citep{shemilt_2022_surface} is amplified by introducing surfactant. Using numerical solutions of the long-wave equations, which describe the motion of thick films, we showed that the approximately four-fold increase in plugging time when strong surfactant is introduced to a thick Newtonian layer \citep{otis_effect_1990,halpern_surfactant_1993} can become much larger when the liquid is viscoplastic (figure \ref{fig:LWsimulation}). The critical layer thickness required for a plug to form is also known to increase as the capillary Bingham number is increased \citep{shemilt_2022_surface}. In figure \ref{fig:LWsweep}, we quantified how this increase is amplified when sufficiently strong surfactant is present; the threshold for surfactant to have an appreciable effect corresponds to a relative surface-tension reduction of only a few percent. By examining the limit of large Marangoni number in \S \ref{sec:LWlargeM}, we found that surface velocities are induced by local changes to the shape of the interface during plug formation, which act to redistribute surfactant and uniformly raise the concentration globally. 

The results suggest a mechanism for the stabilising effect of pulmonary surfactant in the small airways, particularly in diseases where mucus yield stress is increased such as cystic fibrosis \citep{patarin_rheological_2020}. Figure \ref{fig:LWsimulation}(b) illustrates the delay to plug formation caused by surfactant. In these simulations, plug formation occurs at $t_p\approx 35$ when there is no surfactant, and $t_p\approx268$ when there is surfactant. We can relate these dimensionless times to real time scales in the lungs by taking the following physiologically relevant parameters values for a $12^{th}$ generation airway: airway radius, $a=0.4$mm \citep{terjung_lung_2016}, equilibrium surface tension, $\sigma_0=30$mN/m \citep{chen_determination_2019} and viscosity, $\eta=0.01$Pas \citep{lai_micro-_2009}. Using these values, $t_p\approx35$ and $t_p\approx268$ correspond to real plugging times of $t_p^*\approx1.7$ secs and $t_p^*\approx13$ secs. Whilst there is significant variation in measurements of rheological parameters for mucus, this suggests that for a $12^{th}$ generation airway, the plugging time in the surfactant-free simulation is less than the time of a breathing cycle and it is longer than a breathing cycle in the surfactant-laden simulation. This provides evidence that surfactant can make plug formation less likely to occur in an airway. 

The results presented in figure \ref{fig:LWsweep} for the critical layer thickness for plugging to occur also have physiological relevance. Measurements of healthy pulmonary surfactant suggest that Marangoni numbers around $\Ma\approx1$ are likely to be observed \citep{SCHURCH2001195}, although there is significant variation in measured values. 
Figure \ref{fig:LWsweep}(b) illustrates how the critical thickness is increased when $\Ma\approx1$ compared to when $\Ma$ is low. There is clinical evidence that in lungs affected by cystic fibrosis, surfactant function is impaired \citep{pulmonary_2017_gunasekara} and that this is linked to increased prevalence of airway obstructions \citep{griese_2004_pulmonary}. Our results are consistent with these clinical observations, and suggest a mechanism for how surfactant deficiency may destabilise airways. Moreover, as discussed in \citet{shemilt_2022_surface}, mucolytic therapies commonly used in cystic fibrosis can significantly decrease mucus yield stress \citep{patarin_rheological_2020}, which could destabilise airways and trigger plug formation if the mucus layer is thick. One of these mucolytic drugs, rhDNAse, has been linked with increased lung exacerbations and a decline in lung function in patients with idiopathic bronchiectasis \citep{ODONNELL19981329}. The evidence of surfactant amplifying yield-stress effects suggests surfactant needs to be accounted for when considering the impacts of mucolytic drugs that significantly lower the mucus yield stress.

Our results also suggest that introducing surfactant can decrease the peak value of the shear stress induced on the tube wall during the evolution of the liquid layer (figure \ref{fig:TFevolution}c). This suggests that surfactants in airways are likely to provide protection against epithelial cell damage, which can occur when a sufficiently large stress is exerted on the airway wall \citep{huh_acoustically_2007}. This is consistent with previously reported effects of introducing surfactant to a Newtonian layer \citep{romano_2022_surfactant}.

The theory we have used here could be easily modified to study other related flows. For example, the surface Marangoni force could be replaced by a force caused by an oscillating air flow, something that has been studied previously in the case that the liquid is Newtonian \citep{halpern_nonlinear_2003}. Beyond the application to airway modelling, the present theory could be used to study the effect of surfactant on other capillary flows of viscoplastic fluids. For example, one possible industrial application that has received recent modelling attention is ink-jet printing \citep{jalaal_stoeber_balmforth_2021,van_der_kolk_tieman_jalaal_2023}, where the effects of surfactant have not yet been studied but may be important. The flow region maps that we have introduced (figures \ref{fig:yieldmap}, \ref{fig:TFevolution}(a) and \ref{fig:smallMasim}) may also prove useful in other problems by illustrating the transitions between different types of yielding. 

There are many limitations to the modelling approach we have taken. Several physiologically relevant effects have not been included in the model, such as gravity, air flow, and more complex liquid rheologies. However, the relative simplicity of the model has allowed for extensive exploration of parameter space, in order to systematically examine the effect of surfactant on this flow. We have assumed a linear relation between surface tension and surfactant concentration, despite this typically being nonlinear for pulmonary surfactants \citep{SCHURCH2001195}. With this assumption, the model is simplified but still retains sufficient complexity to capture the key effects of surfactant on the flow. As discussed in \S\ref{sec:LWlargeM}, whilst we have explored a wide range of $\Ma$, the linear surfactant equation of state limits how large we can make $\Ma$ in our long-wave simulations. To fully explore the large-$\Ma$ limit in the future, a nonlinear equation of state should be used. The quasi-one-dimensional long-wave theory has been shown to provide a good approximation to the dynamics of thick films in similar problems \citep[e.g., ][]{halpern_effect_2010} so we expect it to perform well here also, but our predictions await validation against fully two-dimensional CFD simulations. Long-wave theory is known to break down immediately before plug formation \citep{johnson_nonlinear_1991}, so to extend the analysis to study coalescence and the post-coalescence phase would require a fully two-dimensional model, in line with what has been done in the Newtonian case \citep{romano_2022_surfactant}. 

In this study, we have quantified how surfactant can amplify the effect of viscoplastic rheology during the capillary instability of a liquid film coating a tube. In addition to slowing the dynamics, sufficiently strong Marangoni forces can induce an effective doubling of the capillary Bingham number for thin films. For thick films with a large enough capillary Bingham number, the critical thickness required for plug formation to occur can be significantly increased by the presence of surfactant. These results suggest a mechanism for how pulmonary surfactant can stabilise airways, and how surfactant deficiency can contribute to the prevalence of mucus plugging in obstructive lung diseases.

\backsection[Acknowledgements]{AH, OEJ and CAW are all supported by the NIHR Manchester Biomedical Research Centre. The views expressed in this publication are those of the author(s) and not necessarily those of the NHS, the National Institute for Health Research, Health Education England or the Department of Health. }

\backsection[Funding]{JDS was supported by an Engineering and Physical Sciences Research Council Doctoral Training Award. This research was co-funded by the NIHR Manchester Biomedical Research Centre (A.H., grant number NIHR203308) and the Engineering and Physical Sciences Research Council (A.B.T., grant number EP/T021365/1).}

\backsection[Declaration of interests]{The authors report no conflict of interest.}

\backsection[Data availability statement]{The code used to generate the data in this study is openly available at https://doi.org/10.5281/zenodo.7794532.}

\backsection[Author ORCID]{J.D. Shemilt, https://orcid.org/0000-0002-9158-0930; A. Horsley, https://orcid.org/0000-0003-1828-0058; O.E. Jensen,  https://orcid.org/
0000-0003-0172-6578; A.B. Thompson,  https://orcid.org/
0000-0001-9558-1554; C.A. Whitfield,  https://orcid.org/
0000-0001-5913-735X}

\appendix 

\section{Derivation of long-wave evolution equations\label{app:LWderivation}}


To derive evolution equations in the long-wave limit, we insert the scaled variables \eqref{LWscaling} into the dimensionless governing equations \eqref{masscons}-\eqref{surftransport1} and then truncate at leading order in $\delta$. Mass and momentum conservation, to leading order, are
\refstepcounter{equation}
$$
    0 = \partial_{\Bar{z}}\bar{w} +\frac{1}{r}\partial_r(r\Bar{u}),\quad0=\partial_r{p},\quad
    \partial_{\Bar{z}}{p}=\frac{1}{r}\partial_r(r\bar{\tau}_{rz}).
    \eqno{(\theequation{\mathit{a}-{c}})}\label{Amomeqns}
$$
The no slip boundary conditions at the wall are
\begin{equation}
    \Bar{u} = \bar{w} = 0 \quad \mbox{on} \quad r= 1.
    \label{AwallBCsLW}
\end{equation}
and the interfacial boundary conditions are
\refstepcounter{equation}
$$  
    \partial_{\bar{t}}R+\bar{w}\partial_{\bar{z}}R = \bar{u},\quad{p} = -\kappa [1+\Ma(1-\Gamma)], \quad\bar\tau_{rz} = {\Ma}\partial_{\bar{z}}\Gamma \quad\mbox{on}\quad r = R.
    \eqno{(\theequation{\mathit{a-c}})}\label{ALWBCs}
$$
As discussed in \S\ref{sec:LWmethods}, in (\ref{ALWBCs}b) we retain the full expression for $\kappa$ \eqref{kappa} rather than truncating it. The surfactant transport equation \eqref{surftransport1} at leading order is
\begin{equation}
    \partial_{\bar{t}}\left(R\Gamma\right) + \partial_{\bar{z}}\left(\bar{w}_sR\Gamma\right) = 0,
    \label{AsurftransportLW}
\end{equation}
where $\bar{w}_s$ is the leading-order surface velocity. Using (\ref{Amomeqns}a), (\ref{AwallBCsLW}) and (\ref{ALWBCs}a), we can derive the evolution equation,
\begin{equation}
    \p_{\bar{t}}R = \frac{1}{R}\p_{\bar{z}}\bar{Q},\quad\mbox{where}\quad \bar{Q} = \int_R^1\bar{w}r\,\mathrm{d}r.
    \label{Aevoleqn}
\end{equation}
The equations \eqref{AsurftransportLW} and \eqref{Aevoleqn} are the long-wave evolution equations for $R$ and $\Gamma$, but they need to be closed by deriving expressions for the surface velocity, $\bar{w}_s$, and the axial volume flux, $\bar{Q}$, which we pursue below. When presenting the equations \eqref{AsurftransportLW} and \eqref{Aevoleqn} in \S\ref{sec:LWmethods}, we rewrite them in terms of the unscaled variables \eqref{nondim}, but they are entirely equivalent to the versions given here in terms of the scaled variables. 

\begin{figure}
    \centering
    \includegraphics[width = 0.7\textwidth]{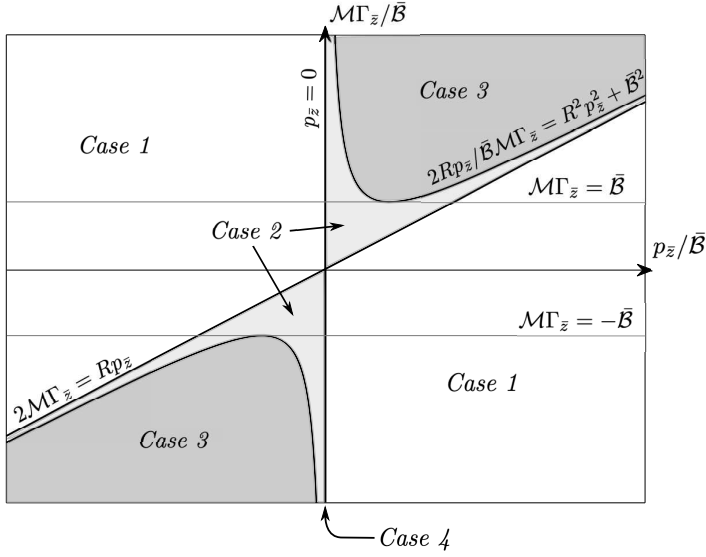}
    \caption{Map of $(p_{\bar{z}}/\bar{\B},\Ma\Gamma_{\bar{z}}/\bar{\B})$-space with the shading indicating the locations of Cases 1-4, which are considered separately to derive the evolution equations in Appendix \ref{app:LWderivation}. In this figure, subscripts denote derivatives. The map shown here, corresponds exactly with figure \ref{fig:yieldmap}(b), which shows the location of the yielding types I-V. Unlike in figure \ref{fig:yieldmap}(b), here the map is plotted in terms of scaled variables \eqref{LWscaling}, but this just corresponds to a uniform scaling of the whole space by $\delta$.}
    \label{fig:Acasemap}
\end{figure}

From (\ref{Amomeqns}b) and (\ref{ALWBCs}b), we deduce that
\begin{equation}
    p(\bar{z},\bar{t}) = -\kappa\left[1+\mathcal{M}(1-\Gamma)\right].
\end{equation}
Integrating (\ref{Amomeqns}c) and using (\ref{ALWBCs}c), we get an expression for the leading-order shear stress,
\begin{equation}
    \bar\tau_{rz} = \frac{\p_{\bar{z}}p}{2}\left(r-\frac{R^2}{r}\right) + \frac{R}{r}{\Ma}\p_{\bar{z}}\Gamma.\label{Ataurz}
\end{equation}
Note that \eqref{Ataurz} holds independently of any rheological considerations. The non-zero components of the strain-rate tensor, up to $O(\delta^2)$, are
\refstepcounter{equation}
$$
    \dot\gamma_{rz} \sim \delta\p_r\bar{w},\quad \dot\gamma_{zz} \sim 2\delta^2\p_{\bar{z}}\bar{w},\quad\dot\gamma_{rr} \sim 2\delta^2\p_r{\bar{u}},\quad \dot\gamma_{\theta\theta} \sim \delta^2\frac{\bar{u}}{r}.
    \eqno{(\theequation{\mathit{a-d}})}
$$
Therefore, the constitutive relation \eqref{constitnodim} implies that, if $|\bar{\tau}_{rz}|> \bar{\mathcal{B}}$, where $\bar{\mathcal{B}} = \mathcal{B}/\delta$, then 
\begin{equation}
    \bar{\tau}_{rz} = \left(1+\frac{\bar{\mathcal{B}}}{|\p_{r}\bar{w}|}\right)\p_{r}\bar{w},
    \label{Aconstit}
\end{equation}
and the leading-order normal stresses are all of size $O(\delta)$. We call regions of the flow where $|\bar{\tau}_{rz}|>\bar{\mathcal{B}}$ shear-dominated, or fully-yielded. As identified by \citet{BALMFORTH199965} for thin-film flows, regions where $|\bar{\tau}_{rz}|\leq\bar{\mathcal{B}}$ exhibit plug-like flow, with $\p_r\bar{w}=0$ to leading order. These plug-like regions can be yielded, with the normal stresses becoming $O(1)$ and the second invariant of the stress becoming exactly equal to $\bar{\mathcal{B}}$ to leading order. To derive an expression for $\bar{w}$ that holds everywhere in the layer, we will use \eqref{Ataurz} to determine the boundaries between the shear-dominated and plug-like regions, which are the surfaces satisfying $|\bar{\tau}_{rz}|=\bar{\mathcal{B}}$, then use \eqref{Aconstit} to find $\bar{w}$ in the shear-dominated regions and, finally, match these to the $r$-independent expression for $\bar{w}$ in the plug-like regions. Note that since $\bar{\tau}_{rz}$ is quadratic in $r$ \eqref{Ataurz}, there can be at most two shear-dominated regions and one plug-like region within $R\leq r\leq1$. We will define the boundaries between shear-dominated and plug-like regions as $r=\Psi_\pm$, such that the shear-dominated regions are $R\leq r\leq\Psi_-$ and $\Psi_+\leq r\leq1$ and the plug-like region is $\Psi_-\leq r\leq\Psi_+$. Any of these three regions may not exist at a given point, in which case the width of it will be zero. There are four separate cases that we must consider separately below, depending on how many roots there are to each of the equations $\bar{\tau}_{rz}=\pm\bar{\B}$. Figure \ref{fig:Acasemap} illustrates where in $(p_{\bar{z}}/\bar{\B},\Ma\Gamma_{\bar{z}}/\bar{\B})$-space each of these four cases occurs. 

\textit{Case 1:} $2{\Ma}\p_{\bar{z}}\Gamma/(R\p_{\bar{z}}p)<1$ and $\p_{\bar{z}}p\neq0$. Figure \ref{fig:Acasemap} shows that this case occurs in the upper-left and lower-right regions of $(p_{\bar{z}}/\bar{\B},\Ma\Gamma_{\bar{z}}/\bar{\B})$-space, where capillary and Marangoni forces either act in opposite directions or, if they act in the same direction, the Marangoni force is relatively weak. 

In this case, $\bar{\tau}_{rz}$ is monotonic for $r>0$, so there is exactly one solution to $\bar{\tau}_{rz}=\bar{\B}$ and exactly one solution to $\bar{\tau}_{rz}=-\bar{\B}$. We define $r=\psi_{\pm}^{(1)}$ as the surfaces such that $\bar{\tau}_{rz}=\pm \bar{\B}\sgn(\p_{\bar{z}}p)$. From \eqref{Ataurz}, we get
\begin{equation}
    \psi_{\pm}^{(1)} = \pm\frac{\bar{\B}}{|\p_{\bar{z}}{p}|}+\sqrt{\left(\frac{\bar{\B}}{\p_{\bar{z}}{p}}\right)^2 + R^2 - \frac{2R{\Ma}\p_{\bar{z}}\Gamma}{\p_{\bar{z}}{p}}}.
    \label{Apsipm1}
\end{equation}
The functions $\psi_{\pm}^{(1)}$ could take any positive value, and whether the surfaces $r=\psi_{\pm}^{(1)}$ lie within the layer, $R<r<1$, or not determines which of the shear-dominated and plug-like regions exist. If both surfaces lie within the layer, i.e. $R<\psi_-^{(1)}\leq\psi_+^{(1)}<1$, then $\Psi_{\pm}=\psi_{\pm}^{(1)}$ and both shear-dominated regions and the plug-like region exist. We say that the layer is exhibiting yielding of type I (see figure \ref{fig:yieldmap}a). If, instead, $\psi_-^{(1)}\leq R<\psi_+^{(1)}<1$ then $\Psi_+ = \psi_+^{(1)}$ and $\Psi_-=R$, and there is yielding of type II. If $\psi_\pm^{(1)}\leq R$ then $\Psi_\pm=R$ and there is yielding of type III. Similarly, if $\psi_{\pm}^{(1)}\geq1$ then $\Psi_{\pm}=1$ and there is also yielding of type III. If $R<\psi_-^{(1)}<1\leq\psi_+^{(1)}$, then $\Psi_-=\psi_-^{(1)}$ and $\Psi_+=1$, and there is yielding of type IV. Finally, if $\psi_-^{(1)}<R<1<\psi^{(1)}_+$ then $\Psi_-=R$, $\Psi_+=1$ and there is no yielding (type V). We can identify that all of the above cases are captured by defining
\begin{equation}
\Psi_\pm=\max\left[R,\min\left(1,\psi_{\pm}^{(1)}\right)\right].
\end{equation}

With the expressions for $\Psi_\pm$ now established, we can proceed to determine $\bar{w}$. In $\Psi_+\leq r \leq1$, we equate \eqref{Ataurz} with \eqref{Aconstit}, solve for $\p_{r}\bar{w}$, then integrate and use \eqref{AwallBCsLW}. We also need to use the fact that $\sgn(\p_r\bar{w})=\sgn(\p_{\bar{z}}p)$ in $\Psi_+\leq r \leq1$. Doing this gives
\begin{equation}
    \bar{w} = 
\frac{\p_{\bar{z}}{p}}{4}\left(r^2-1-2R^2\log r\right)+{\Ma}(\p_{\bar{z}}\Gamma) R\log r -{\bar{\B}}\sgn(\p_{\bar{z}}{p})(r-1)
\label{Awplus1}
\end{equation}
for $\Psi_+\leq r \leq1$. We then determine the velocity in the plug-like region, $\Psi_-\leq r\leq\Psi_+$, by evaluating \eqref{Awplus1} at $r=\Psi_+$. Therefore, $\bar{w}=\bar{w}_p(\bar{z},\bar{t})$ in $\Psi_-\leq r\leq\Psi_+$, where
\begin{equation}
    \bar{w}_p=\frac{\p_{\bar{z}}{p}}{4}\left(\Psi_+^2-1-2R^2\log\Psi_+\right)
+{\Ma}(\p_{\bar{z}}\Gamma)R\log\Psi_+-{\bar{\B}}\sgn(\p_{\bar{z}}{p})(\Psi_+-1).
\label{Awp1}
\end{equation}
From \eqref{Awp1}, we have the value of $\bar{w}$ at $r=\Psi_-$, which can be used to find $\bar{w}$ in the shear-dominated region, $R\leq r\leq\Psi_-$, using the same procedure as above, but noting that instead $\sgn(\p_r\bar{w})=-\sgn(\p_{\bar{z}}p)$. We get
\begin{equation}
    \bar{w} = \bar{w}_p + \frac{\p_{\bar{z}}{p}}{4}\left[r^2-\Psi_-^2-2R^2\log\left(\frac{r}{\Psi_-}\right)\right]+{\Ma}(\p_{\bar{z}}\Gamma)R\log\left(\frac{r}{\Psi_-}\right)+\bar{\B}\sgn(\p_{\bar{z}}{p})(r-\Psi_-)
    \label{Awmin1}
\end{equation}
for $R\leq r\leq\Psi_-$. This then completes the leading-order axial velocity in Case 1. We can then integrate to determine volume flux, $\bar{Q}$, and evaluate \eqref{Awmin1} at $r=R$ to get the surface velocity, $\bar{w}_s$. This gives
\begin{equation}
    \bar{Q} = -\frac{\p_{\bar{z}}p}{16}F_1 - \frac{1}{4}\Ma(\p_{\bar{z}}\Gamma)R F_2 - \frac{\bar{\B}}{6}\sgn(\p_{\bar{z}}p)(F_3+F_4),
    \label{AQ1}
\end{equation}
where the functions $F_1$, $F_2$, $F_3$ and $F_4$ are given in \eqref{F1}, and
\begin{equation}
    \bar{w}_s = \frac{\p_{\bar{z}}p}{4}G_1 + \Ma(\p_{\bar{z}}\Gamma)RG_2 + \bar{\B}\sgn(\p_{\bar{z}}p)(G_3+G_4),
    \label{Aws1}
\end{equation}
where the functions $G_1$, $G_2$, $G_3$ and $G_4$ are given in \eqref{G1}. 

\textit{Case 2:} $1 + {\bar{\B}}^2/(R\p_{\bar{z}}p)^2\geq{2{\Ma}\p_{\bar{z}}\Gamma}/({R\p_{\bar{z}}p})\geq1$ and $\p_{\bar{z}}p\neq0$. Again, figure \ref{fig:Acasemap} illustrates the region of $(p_{\bar{z}}/\bar{\B},\Ma\Gamma_{\bar{z}}/\bar{\B})$-space where this case occurs. In this case, there are no solutions to $\bar{\tau}_{rz}=-\bar{\B}\sgn(\p_{\bar{z}}p)$ in $r\geq0$. We define $r=\psi_{\pm}^{(2)}$ as the two surfaces on which $\bar{\tau}_{rz}=\bar{\B}\sgn(\p_{\bar{z}}p)$, with $\psi_{-}^{(2)}\leq\psi_{+}^{(2)}$. From \eqref{Ataurz}, these are
\begin{equation}
    \psi_{\pm}^{(2)} = \frac{\bar{\B}}{|\p_{\bar{z}}p|}\pm\sqrt{\left(\frac{\bar{\B}}{\p_{\bar{z}}p}\right)^2 + R^2 - \frac{2R{M}\p_{\bar{z}}\Gamma}{\p_{\bar{z}}p}}.
    \label{psipm2}
\end{equation}
As in Case 1 above, by considering all possible types of yielding that can occur, we find
\begin{equation}
\Psi_\pm= \max\left[R,\min\left(1,\psi_{\pm}^{(2)}\right)\right].
\end{equation}
We can then proceed similarly to above to derive $\bar{w}$. The only difference here is that $\p_r\bar{w}$ has the opposite sign in the near-interface shear-dominated region, $R\leq r\leq\Psi_-$, compared to Case 1. Hence, $\bar{w}$ is still given by \eqref{Awplus1} in $\Psi_+\leq r\leq1$, and by \eqref{Awp1} in $\Psi_-\leq r \leq\Psi_+$. However, we have
\begin{equation}
    \bar{w} = \bar{w}_p + \frac{\p_{\bar{z}}{p}}{4}\left[r^2-\Psi_-^2-2R^2\log\left(\frac{r}{\Psi_-}\right)\right]+{\Ma}(\p_{\bar{z}}\Gamma)R\log\left(\frac{r}{\Psi_-}\right)-\bar{\B}\sgn(\p_{\bar{z}}{p})(r-\Psi_-)
    \label{Awmin2}
\end{equation}
in $R\leq r \leq\Psi_-$. This leads to slightly modified expressions for axial volume flux,
\begin{equation}
        \bar{Q} = -\frac{\p_{\bar{z}}p}{16}F_1 - \frac{1}{4}\Ma(\p_{\bar{z}}\Gamma)R F_2 - \frac{\bar{\B}}{6}\sgn(\p_{\bar{z}}p)(F_3-F_4),
    \label{AQ2}
\end{equation}
and surface velocity,
\begin{equation}
        \bar{w}_s = \frac{\p_{\bar{z}}p}{4}G_1 + \Ma(\p_{\bar{z}}\Gamma)RG_2 + \bar{\B}\sgn(\p_{\bar{z}}p)(G_3-G_4),
    \label{Aws2}
\end{equation}
in Case 2.

\textit{Case 3}: $2{\Ma}\p_{\bar{z}}\Gamma/(R\p_{\bar{z}}{p})>1+ \bar{{\B}}^2/({R\p_{\bar{z}}p})^2$. Figure \ref{fig:Acasemap} shows that, in this case, capillary and Marangoni forces act in the same direction, with Marangoni forces being relatively strong. In Case 3, $|\bar{\tau}_{rz}|>\bar{\B}$ for all $r>0$. Hence, the whole layer is shear-dominated (yield type III). We set $\Psi_+=\Psi_-=R$. The axial velocity, $\bar{w}$ is then given by \eqref{Awplus1} for the whole layer, $R\leq r \leq 1$. Since there is no contribution from the plug-like region or the near-interface shear-dominated region, the expressions for $\bar{Q}$ and $\bar{w}_s$ from Case 1, \eqref{AQ1} and \eqref{Aws1}, and from Case 2, \eqref{AQ2} and \eqref{Aws2}, both recover the correct expressions for $\bar{Q}$ and $\bar{w}_s$ in Case 3, so either can be used. 

\textit{Case 4:} $\p_{\bar{z}}p=0$. When $\p_{\bar{z}}p=0$, there is exactly one solution to $\bar{\tau}_{rz}=\bar{\B}\sgn(\p_{\bar{z}}\Gamma)$ and no other solutions to $|\bar{\tau}_{rz}|=\bar{\B}$ in $r>0$. The solution corresponds to $r=\Psi_-$. We can deduce from \eqref{Ataurz} that $\Psi_-=\min(1,\max[R,R\Ma|\Gamma_z|/\bar{\B}])$, and we always have $\Psi_+=1$. This is consistent with the behaviour in Cases 1 and 2 as $\p_{\bar{z}}p\rightarrow0$. The velocity, $\bar{w}$, can be derived using a similar procedure as in the cases above: we equate \eqref{Ataurz} and \eqref{Aconstit} in $R\leq r\leq\Psi_-$, solve for $\p_r{\bar{w}}$, then integrate and apply no slip at $r=\Psi_-$ to get $\bar{w}$. We have $\bar{w}=0$ in $\Psi_-\leq r \leq1$. Again, we integrate the velocity to get the flux,
\begin{equation}
    \bar{Q} = -\frac{1}{4}R\Ma(\p_{\bar{z}}\Gamma)F_2 + \bar{\B}\sgn(\p_{\bar{z}}\Gamma)F_4,
\end{equation}
and evaluate $\bar{w}$ at $r=R$ to get the surface velocity,
\begin{equation}
    \bar{w}_s = \Ma(\p_{\bar{z}}\Gamma)RG_2 - \bar{\B}\sgn(\p_{\bar{z}}\Gamma)G_4,
\end{equation}
in Case 4. 

We have now derived expressions for the axial volume flux and surface velocity in all cases, and so have closed the evolution equations \eqref{AsurftransportLW} and \eqref{Aevoleqn}. Finally, the lateral boundary conditions \eqref{sideBCs1}, at leading order in $\delta$, imply
\begin{equation}
    \p_{\bar{z}}R = \bar{Q} = \bar{w}_s\Gamma \quad\mbox{on}\quad \bar{z} = \bar{L}.
\end{equation}
Equations \eqref{LWevoleqn}-\eqref{LWsideBCs}, are presented in terms of the unscaled variables \eqref{nondim}, but they are entirely equivalent to what is derived here.

\section{Rankine-Hugoniot condition for shock propagation speed}\label{app:propagation}

Suppose we observe a jump discontinuity in $\Gamma_z$ at the point $z=z_s(t)$, and denote the size of the jump by $[\Gamma_z]^+_-$. Then we can use the following argument to derive a Rankine-Hugoniot condition (see, e.g., \cite{billingham_king_2001}) for the speed of propagation of the discontinuity. 
Define two points, $z_1$ and $z_2$, such that $0<z_1<z_s(t)<z_2<L$. Then the surfactant transport equation \eqref{surftransportTF} implies
\begin{equation}
    \frac{\mathrm{d}}{\mathrm{d}t}\int_{z_1}^{z_2}\Gamma_z\,\mathrm{d}z= -\left[\left(\tilde{w}_s\Gamma\right)_z\right]_{z_1}^{z_2}. \label{appRheqn1}
\end{equation}
Splitting the integral in \eqref{appRheqn1} and expanding, we get
\begin{equation}
    \frac{\mathrm{d}}{\mathrm{d}t}\left(\int_{z_1}^{z_s}\Gamma_z\,\mathrm{d}z+\int_{z_s}^{z_2}\Gamma_z\,\mathrm{d}z\right) = -\frac{\mathrm{d}z_s}{\mathrm{d}t}\left[\Gamma_z\right]_{z_1}^{z_2} + \int_{z_1}^{z_s}\Gamma_{tz}\,\mathrm{d}z+\int_{z_s}^{z_2}\Gamma_{tz}\,\mathrm{d}z.\label{appRheqn2}
\end{equation}
Then taking $z_1\rightarrow z_s^-$ and $z_2\rightarrow z_s^+$, \eqref{appRheqn1} and \eqref{appRheqn2} imply
\begin{equation}
    \frac{\mathrm{d}z_s}{\mathrm{d}t}\left[\Gamma_z\right]_{-}^{+} = \left[\left(\tilde{w}_s\Gamma\right)_z\right]_{-}^{+},
\end{equation}
which defines the shock propagation speed
\begin{equation}
    u_s \equiv \frac{\mathrm{d}z_s}{\mathrm{d}t} = \frac{\left[\left(\tilde{w}_s\Gamma\right)_z\right]_{-}^{+}}{\left[\Gamma_z\right]_{-}^{+}}.\label{appRHcondition}
\end{equation}
\color{black}

The velocity \eqref{appRHcondition} can, in theory, be integrated to get the shock location,
\begin{equation}
    z_s(t) = z_s(t_0) + \int_{t_0}^tu_s(t')\mathrm{d}t',\label{appRHzs}
\end{equation}
if the location of the shock is known at some time, $t_0$. We have used \eqref{appRHzs} to provide a consistency check on numerical simulations. For example, for the simulation shown in figure \ref{fig:TFevolution}(a), we take $t_0=69.2$, which is shortly after the interface-adjacent yielded region first appears, we take the value of $z_s(t_0)$ from the simulation and we use the values for $[\tilde{w}_s\Gamma_z]_-^+$ and $[\Gamma_z]_-^+$ also from the simulation. Doing this, we find that the prediction of the location of either shock computed via \eqref{appRHzs} never deviates from the shock location in the simulation by more than three grid points (when using $200$ points), providing evidence that the numerical scheme is capturing the speed of shock propagation. Since we can only calculate $u_s$ by using values from the numerical simulation this does not provide independent validation of the numerical method. However, it does provide evidence that at each instant the shock propagation velocity is being accurately calculated in the numerical scheme
from the finite differenced derivatives and that spurious behaviour is not being introduced by the numerical scheme.

\color{black}

\section{Thin-film dynamics at large $M$ \label{app:TFlargeM}}

Consider the thin-film equations \eqref{TFevoleqn}-\eqref{TFsideBCs} in the limit of very strong surfactant. We propose the expansions
\begin{equation}
    \left.\begin{array}{c}
    H = \mathcal{H}_0 + \frac{1}{M}\mathcal{H}_1+\dots,\quad \tilde{p} = \tilde{p}_0 + \frac{1}{M}\tilde{p}_1+\dots,\quad \tilde{w}_s = \tilde{w}_0 + \frac{1}{M}\tilde{w}_1 + \dots,\vspace{4pt}\\
    \Gamma = \tilde{\mathcal{G}}_0 + \frac{1}{M}\tilde{\mathcal{G}}_1+\dots,\quad
    Y_- = {Y}_{-0} + \frac{1}{M}{Y}_{-1}+\dots,\quad Y_+ = {Y}_{+0} + \frac{1}{M}{Y}_{+1}+\dots,
    \end{array}\right\}\label{largeMexpansions}
\end{equation}
as $M\rightarrow\infty$. The Marangoni force at the interface must be finite in the limit $M\rightarrow\infty$, so we require $|M\Gamma_z|<\infty$. This implies $\tilde{\mathcal{G}}_{0,z}=0$ and conservation of total mass of surfactant then means we must have 
\begin{equation}
    \tilde{\mathcal{G}}_0=1.\label{CG0}
\end{equation}
Inserting \eqref{largeMexpansions} into the surfactant transport equation \eqref{surftransportTF}, and using \eqref{CG0}, implies $\tilde{w}_{0,z}=0$. Combined with the boundary condition \eqref{TFsideBCs} which enforces $\tilde{w}_0\tilde{\mathcal{G}}_0=0$ at $z=\{0,L\}$, we get
\begin{equation}
    \tilde{w}_0=0.\label{Cw0}
\end{equation}
This means that in the limit of strong surfactant, the interface of the thin film is essentially immobilised by Marangoni effects. Inserting \eqref{Cw0} into the expression for surface velocity \eqref{wsTF}, and rearranging gives
\begin{equation}
    \mathcal{H}_0 - Y_{+0} = Y_{-0},\label{CHYY}
\end{equation}
assuming $\tilde{p}_{0,z}\neq0$. The result \eqref{CHYY} says that the wall-adjacent and interface-adjacent fully yielded regions always have the same thickness in the leading-order theory. This also means that only yielding of types I or V can occur. We will assume for now that the fluid is yielded, so $Y_->0$ and $Y_+<H$, and then subsequently present criteria for when the fluid rigidifies. 

The definition \eqref{Ydefns} gives
\begin{equation}
    {Y}_{\pm0} = \mathcal{H}_0 \pm \frac{B}{|\tilde{p}_{0,z}|} + \frac{\tilde{\mathcal{G}}_{1,z}}{\tilde{p}_{0,z}},
\end{equation}
which when combined with \eqref{CHYY}, implies
\begin{equation}
    \tilde{\mathcal{G}}_{1,z} + \frac{1}{2}\mathcal{H}_0\tilde{p}_{0,z}=0.\label{CG1z}
\end{equation}
Note that, for certain choices of initial conditions, $\tilde{\mathcal{G}}_1$ may not initially satisfy \eqref{CG1z}, in which case we expect there to be a rapid adjustment at early times to a state where \eqref{CG1z} is satisfied. For an arbitrary choice of initial $\Gamma$, the change in $\Gamma$ required to reach a state satisfying \eqref{CG1z} is of size $O(1/M)$, so we expect the time scale for the adjustment to this state to be of size $O(1/M)$ also, since the initial surface velocity would be of size $O(1)$ during the adjustment period. 

At leading order, the axial volume flux \eqref{qdefnTF}, is
\begin{equation}
    q = \frac{1}{6}\tilde{p}_{0,z}Y_{-0}^2\left(2Y_{-0}-3\mathcal{H}\right).\label{Cq}
\end{equation}
If we define the function
\begin{equation}
    \tilde{\mathcal{Y}} \equiv \frac{1}{2}H - \frac{B}{|\tilde{p}_z|},\label{CYtilde}
\end{equation}
then from \eqref{CYtilde}, $Y_{-0}=\tilde{\mathcal{Y}}$ to leading order when the fluid is yielded. Note that, from \eqref{CHYY}, the criterion for no yielding to occur in the leading order theory is $Y_{-0}=0$. Therefore, from \eqref{TFevoleqn} and \eqref{Cq}, the leading order evolution equation, which holds if fluid is yielded or unyielded, is \eqref{largeMevoleqn}.
To reach \eqref{largeMevoleqn}, we have assumed so far that $\tilde{p}_{z}\neq0$, but we can see from \eqref{CG1z} that if $\tilde{p}_{0,z}=0$ then $\tilde{\mathcal{G}}_{1,z}=0$ so there is no driving force. Therefore, \eqref{largeMevoleqn} also holds when $\tilde{p}_z=0$ since it says there is no motion in that case.

\bibliographystyle{jfm}

\bibliography{jfm}

\end{document}